\pdfoutput=1
\documentclass[a4paper,11pt,aps,prd,superscriptaddress,nofootinbib]{revtex4-2}
\usepackage[utf8]{inputenc}
\usepackage[T1]{fontenc}
\usepackage{microtype}
\usepackage{graphicx}
\graphicspath{{./figures/}}

\usepackage{mathtools}
\usepackage{amssymb}
\DeclareMathOperator{\sgn}{sgn}
\DeclareMathOperator{\arccot}{arccot}
\DeclarePairedDelimiter{\abs}{\lvert}{\rvert}

\usepackage[pdfusetitle,bookmarksopen,colorlinks,allcolors=blue]{hyperref}

\begin{document}
\title{Compact kinks in a modified Christ-Lee model}

\author{F.~M.~Hahne}
\email{fernandomhahne@gmail.com}
\affiliation{Programa de Pós-Graduação em Física, Universidade Estadual de Santa Cruz, Campus Soane Nazaré de Andrade, 45662-900, Ilhéus, Brazil}

\author{R.~Thibes}
\email{thibes@uesb.edu.br}
\affiliation{Programa de Pós-Graduação em Física, Universidade Estadual de Santa Cruz, Campus Soane Nazaré de Andrade, 45662-900, Ilhéus, Brazil}
\affiliation{Departamento de Ciências Exatas e Naturais, Universidade Estadual do Sudoeste da Bahia, Rodovia BR 415, km 03 s/n, Itapetinga, Brazil}

\begin{abstract}
    We study compact kinks in a modified Christ-Lee model where the potential is a non-analytic function at the minima. The model has two control parameters that determine the order of the potential and its overall shape. We consider cases in which the potential has a double well shape, a triple well shape, as well as cases with a false vacuum. Compact kink solutions and their internal modes are found for each of these cases. We also study the collision of such kinks and their dependence on the solitons velocity and on the potential parameters. Several interesting dynamical processes are reported, such as the formation of central oscillons for some high velocity collisions, absence of outgoing solitons in cases where the potential is given by piece-wise linear functions, and the decoupling of subkinks from the larger kink in cases with false vacuum, leading to the formation of false vacuum bubbles.
\end{abstract}

\maketitle

\section{Introduction}
\label{sec:introduction}

Topological solitons appear in several areas of physics, including effective models of particle and nuclear physics, condensed matter systems and cosmological settings~\cite{Manton:2004tk,Shnir:2018yzp}.
Specially interesting is the case of scalar fields in $1 + 1$ dimensions, where the presence of multiple vacua is sufficient to allow for the existence of topological solitons called kinks.
Topological kinks and their interactions constitute a present-time hectic research area in mathematical physics~\cite{Kevrekidis:2019zon,Manton:2020onl,Manton:2021ipk,Adam:2021gat,Adam:2023qgx,Pereira:2021gvf,Campos:2024ijb}.
Even in the prototypical double well $\phi^4$ model, several key features of kink collisions -- such as the resonance windows and the value of the critical velocity -- have only recently been explained~\cite{Manton:2020onl,Manton:2021ipk,Adam:2021gat}.
Similar advances followed in other models~\cite{Adam:2023qgx,Pereira:2021gvf,Campos:2024ijb}.

Despite having localized energy density, kinks usually have infinite range, with the field reaching the vacuum only asymptotically through exponential tails.
An interesting exception is the case of compactons, defined as solutions for which the energy density is non-zero only inside a compact region of space.
Compactons have been originally discovered as solutions of a modified Korteweg-de~Vries equation~\cite{Rosenau:1993zz}, however their existence has been demonstrated also in relativistic scalar models when the potential is non-analytic at the minima~\cite{Arodz:2002yt}.

Around the minima, non-analytic potentials have non-zero one-sided derivatives that are different from each side, i.e., they approach the minima with a V-shape.
As a consequence, the field equations for models with a V-shaped potential do not have a linear regime.
This represents a rather dramatic departure from the harmonic oscillator paradigm that permeates theoretical physics.
Non-analytic potentials in field theory have been originally found in the continuum limit of mechanical models with rigid barriers~\cite{Arodz:2005gz}.
However, they have also appeared in the study of the BPS submodels of the Skyrme model, where the V-shaped potential may be obtained from restrictions to field variables due to symmetry reduction~\cite{Adam:2017pdh,Adam:2017srx,Klimas:2018woi}.
Another interesting interpretation of non-analytic potentials is that they can represent hyper-massive limits of usual field theory models~\cite{Bazeia:2014hja,Arodz:2007jh,Hahne:2024qby}.

Compact solutions in models with non-analytic potentials have been the subject of intensive study for the last two decades~\cite{Arodz:2007jh,Bazeia:2014hja,Arodz:2011zm,Swierczynski:2016ivq,Arodz:2005bc,Klimas:2018woi,Hahne:2019ela,Klimas:2023ife,Hahne:2022wyl,Hahne:2023dic,Hahne:2024qby}.
For the case of topological compactons, the interaction between compact kinks with oscillons~\cite{Hahne:2022wyl} and with other kinks~\cite{Hahne:2023dic,Hahne:2024qby} has been the focus of recent studies.
Despite the at first unusual character of the potential, those works have found that the dynamics of compact kinks in models with non-analytic potentials share some characteristics with more usual ones, as the $\phi^4$ model.
For example, the collisions can be broadly categorized into capture and escape cases that alternate in a fractal-like pattern.
Furthermore, their main dynamical features have been recently reproduced by collective coordinates models with the amplitude of the first few kink internal modes as degrees of freedom~\cite{Hahne:2023dic,Hahne:2024qby}.

However, those studies have focused on non-analytic potentials given by piece-wise quadratic expressions, with either periodic or double well shapes.
There has also been some interest in models featuring semi-compact kinks~\cite{Bazeia:2020car}.
Meanwhile, in models with smooth potentials, topological kink interactions have been studied for potentials of higher-order polynomials~\cite{Dorey:2011yw,Weigel:2013kwa,Marjaneh:2017mko,Bazeia:2023qpf,Lima:2024ohp}, as well as models featuring false vacua~\cite{Demirkaya:2017euk,Gomes:2018heu,Dorey:2023izf}.
Indeed, the dynamics of non-compact topological kinks has been scrutinized in a wide range of scenarios, distinguishing model particularities from more general features.
For example, the spectra of internal modes varies from model to model, strongly affecting the resonance windows in kink-antikink collisions.
Similarly, the overall shape of the potential changes the relative size of the core, skin, and tail regions of the kinks, which also affects their interactions~\cite{Karpisek:2024zdj}.
So far, a similarly wide analysis has been missing for the case of compact kinks.

In this work, we study topological kinks in models with non-analytic potentials given by piece-wise expressions of higher power orders, including rational and irrational powers, and with different vacuum structures.
We propose a fairly general bi-parametrized potential, V-shaped precisely at its minima, possessing a rich vacuum structure similar to the Christ-Lee model, including cases of false vacuum.
Besides assuring generality, the two control parameters provide a convenient tool for classification of our analytic and numerical results, easing comparison with previous particular results in the literature.
After obtaining the static kink solutions, we discuss its dynamics, analyzing both its internal modes and dynamics in collision processes.
In this way, we aim to provide a comprehensive and systematic analysis of compact kinks and their interactions in a broader context, so that general features of compactons can be distinguished from specificities of particular models.

This paper is organized as follows.
Inspired by the original Christ-Lee model, in section~\ref{sec:model} we introduce and discuss our bi-parametrized non-analytic potential and classify its vacuum structure.
In section~\ref{sec:static-kinks}, we obtain the static kink solutions calculating their support sizes and masses.
In section~\ref{sec:internal-modes}, we obtain the stability potential and internal modes by numerically solving the corresponding eingenvalue problem.
The simulation results for the kink collisions, along with detailed discussion, is presented in section~\ref{sec:collisions}.
We summarize our results and give some concluding remarks in section~\ref{sec:collisions}.
An appendix with technical details regarding the numerical simulations is also provided.

\section{Model}
\label{sec:model}

We shall construct a non-analytic bi-parametrized self-interacting potential inspired by the standard Christ-Lee model~\cite{Christ:1975wt}, which is a scalar field model in $1+1$ dimensions with Lagrangian density
\begin{equation*}
    \mathcal{L}_{CL} = \frac{1}{2} (\partial_t \phi)^2 - \frac{1}{2} (\partial_x \phi)^2 - V_{CL}(\phi)
\end{equation*}
where the potential function
\begin{equation*}
    V_{CL}(\phi) = \frac{1}{2} \frac{\epsilon^2 + \phi^2}{1 + \epsilon^2} (1 - \phi^2)^2
\end{equation*}
has a parameter $\epsilon$ characterizing its vacuum structure.
For $\epsilon = 0$, the potential $V_{CL}(\phi)$ gives the prototypical $\phi^6$ triple well with three global minima at $\phi \in \{ -1, 0, 1 \}$,
for $ 0 < \epsilon < 1 / \sqrt{2} $ the vacuum at $\phi = 0$ is false,
and for $\epsilon \geq 1 / \sqrt{2}$ the potential only has two minima.
At the $\epsilon \to \infty$ limit, the Christ-Lee model reduces to the well-known $\phi^4$ double well model.
As the Christ-Lee potential approaches each minimum parabolically, its solutions are of infinite range, featuring exponential tails.

\begin{figure}
    \includegraphics{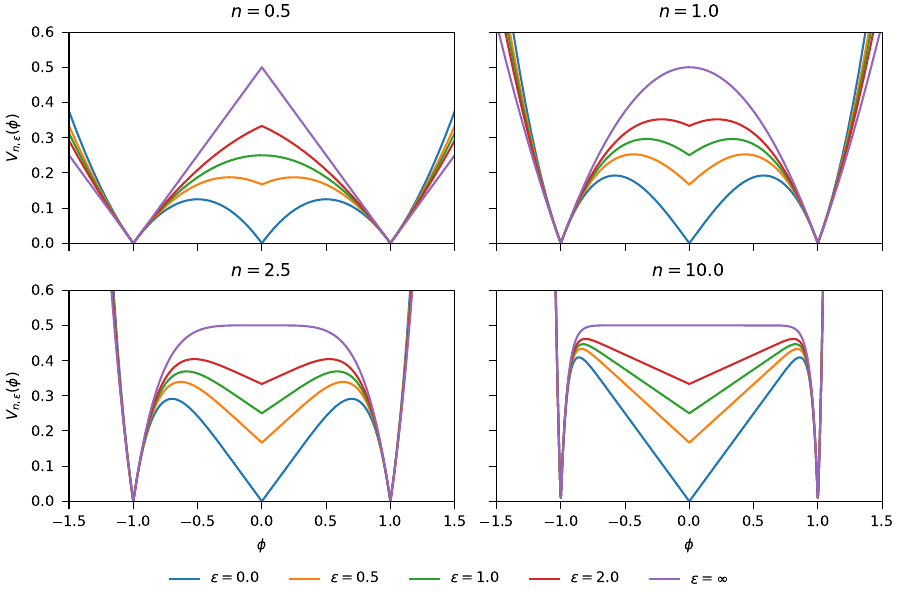}
    \caption{Potential $V_{n,\epsilon}(\phi)$ for different values of the parameters $n$ and $\epsilon$.}
    \label{fig:potential}
\end{figure}

Since our focus is on compact solutions, we propose a modified version of the Christ-Lee model, with the Lagrangian density
\begin{equation*}
    \mathcal{L} = \frac{1}{2} (\partial_t \phi)^2 - \frac{1}{2} (\partial_x \phi)^2 - V_{n,\epsilon}(\phi)
\end{equation*}
containing a bi-parametric non-analytic potential defined as
\begin{equation}
    V_{n,\epsilon}(\phi) = \frac{1}{2} \frac{\epsilon + \abs{\phi}}{1+\epsilon} \abs*{\abs{\phi}^{2n} - 1},
    \label{eq:V}
\end{equation}
where $n \geq 1/2$ and $\epsilon \geq 0$ denote two real parameters.
The presence of the absolute value functions in $V_{n,\epsilon}(\phi)$ makes the potential a non-analytic function.
The parameter $\epsilon$ controls the height of the potential at $\phi = 0$, while $n$ is related to the order of the potential function.
We present some plots of the potential in figure~\ref{fig:potential}.

The field equation is the Euler-Lagrange equation
\begin{equation}
    \partial_t^2 \phi - \partial_x^2 \phi + V'_{n,\epsilon}(\phi) = 0 \label{eq:EL}
\end{equation}
where $V_{n,\epsilon}'(\phi)$ is the potential derivative
\begin{equation}
    V'_{n,\epsilon}(\phi) = \frac{1}{2} \frac{\sgn\phi}{1+\epsilon} \abs*{1-\abs{\phi}^{2n}} + n \frac{\epsilon + \abs{\phi}}{1+\epsilon} \sgn\left(\abs{\phi}^{2n} - 1\right) \sgn(\phi) \abs{\phi}^{2n-1}
    \label{eq:V'}
\end{equation}
in which we impose $\sgn 0 = 0$.
This convention agrees with the definition of derivative in the distributional sense and guarantees that the constant field configuration equal to a minimum of the potential corresponds to a solution of the field equation.
Since the potential has a V-shape around its minima, the solutions of the field equations have compact support, i.e., are different from the vacuum only inside compact regions of space.
From the derivative expression~\eqref{eq:V'}, we see that the condition $n\geq 1/2$ is necessary for the one-sided derivatives at $\phi=0$ to be finite.

The potential in equation~\eqref{eq:V} has global minima at $\phi = \pm 1$, for which $V_{n,\epsilon}(\pm 1) = 0$.
We can check that the potential is indeed V-shaped around these points by calculating its one-sided derivatives
\begin{align*}
    V_{n,\epsilon}'(1^\pm) &= \lim_{\phi \to 1^\pm} V_{n,\epsilon}(\phi) = \pm n, \\
    V_{n,\epsilon}'(-1^\pm) &= \lim_{\phi \to -1^\pm} V_{n,\epsilon}(\phi) = \pm n.
\end{align*}
The potential function also has a critical point at $\phi = 0$, for which
\begin{equation*}
    V_{n,\epsilon}(0) = \frac{\epsilon}{2 + 2\epsilon}.
\end{equation*}
The one-sided derivatives around $\phi=0$ are
\begin{equation*}
    V'_{n,\epsilon}(0^\pm) = \lim_{\phi\to 0^\pm} V'_{n,\epsilon}(\phi) =
    \begin{cases}
        \mp \frac{1}{2} \pm \frac{1}{1+\epsilon} &\text{ if } n = 1/2,\\
        \pm\frac{1}{2 + 2\epsilon} &\text{ if } n > 1/2.
    \end{cases}
\end{equation*}
For $n=1/2$, we have three possibilities:
\begin{itemize}
    \item If $\epsilon = 0$, the potential has a global minimum at $\phi=0$, so the vacua are $\phi \in \{-1, 0, 1\}$.
    \item If $0 < \epsilon < 1$, the potential has a local minimum at $\phi=0$, so the model has two true vacua $\phi = \pm 1$ and a false vacuum $\phi = 0$.
    \item If $\epsilon \geq 1$, the potential has a maximum at $\phi = 0$, and the only vacua are $\phi=\pm1$.
\end{itemize}
This is indeed very similar to the original analytic Christ-Lee potential, which reduces to the triple well $\phi^6$ model when $\epsilon=0$ and has a false vacuum when $0 < \epsilon < 1/\sqrt{2}$.
For $n > 1/2$, the possibilities are:
\begin{itemize}
    \item If $\epsilon = 0$, the potential has a global minimum at $\phi=0$, so the model vacua are $\phi \in \{-1, 0, 1\}$.
    \item If $\epsilon > 0$, the potential has a local minimum at $\phi=0$, so the model has two true vacua at $\phi = \pm 1$ and a false vacuum $\phi = 0$.
\end{itemize}
In the limit $\epsilon \to \infty$ the potential becomes a double well with the expression
\begin{equation*}
    V_{n,\infty}(\phi) = \frac{1}{2} \abs*{\abs{\phi}^{2n} - 1},
\end{equation*}
again quite analogous to the Christ-Lee potential, that becomes the analytic $\phi^4$ model when $\epsilon\to\infty$, and which amounts to the non-analytic limit of the first potential proposed in ref.~\cite{Bazeia:2014hja}, connecting analytic and non-analytic regimes.

\begin{figure}
    \includegraphics{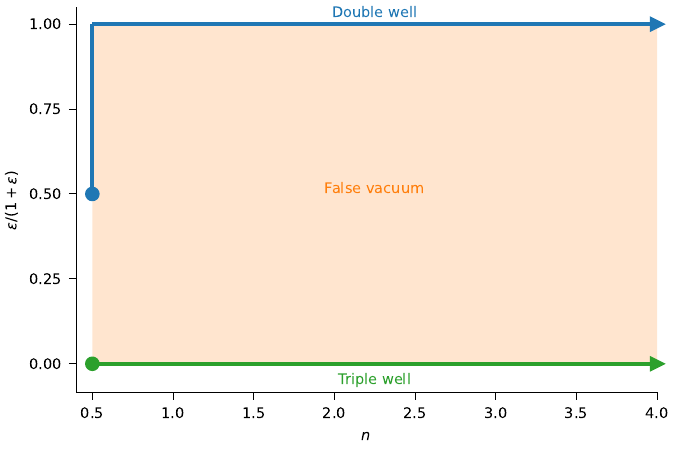}
    \caption{Phase diagram for the different values of parameters $n$ and $\epsilon / (1+ \epsilon)$. We use $\epsilon / (1+ \epsilon)$ instead of $\epsilon$ to be able to represent the case $\epsilon\to \infty$. The blue line represents the regions in the parameter space for which the potential has a double well shape, while the green line represents the region with a triple well potential. For the regions in orange, the potential has a false vacuum. All regions continue towards $n\to \infty$.}
    \label{fig:phase_diagram}
\end{figure}

Summarizing, we have three possible vacuum structures:
\begin{enumerate}
    \item Two true vacua $\phi=\pm1$ and a maxima at $\phi = 0$ for the cases
    \begin{enumerate}
        \item $\epsilon \to \infty$,
        \item $n=1/2$ and $\epsilon \geq 1$.
    \end{enumerate}
    We refer to these cases as the cases with a double well potential.
    \item Three true vacua $\phi \in \{-1, 0, 1\}$ for the case $\epsilon = 0$.
    We refer to these as cases with a triple well potential.
    \item Two true vacua $\phi=\pm 1$ and a false vacuum $\phi = 0$ for the cases
    \begin{enumerate}
        \item $n > 1/2$ and $\epsilon > 0$,
        \item $n = 1/2$ and $0 < \epsilon < 1$.
    \end{enumerate}
    We refer to these as the false vacuum cases.
\end{enumerate}
The different cases can also be represented in a phase diagram, as in figure~\ref{fig:phase_diagram}.

\section{Static kinks}
\label{sec:static-kinks}

The different cases for the vacua structure of the model affect the possible topological solutions.
Topological kinks are possible if $V_{n,\epsilon}(\phi)$ has more than one global minima.
The stability of topological solitons is related to the conservation of the topological charge
\begin{equation*}
    \mathcal{Q} = \int_{-\infty}^{\infty}dx \, \partial_x \phi = \phi(t, x\to\infty) - \phi(t, x\to-\infty),
\end{equation*}
which is determined by the field boundary conditions.
For a given set of boundary conditions, the energy is minimized by static configurations $\phi=\phi(x)$ that solve the Bogomol'nyi–Prasad–Sommerfield (BPS) equation~\cite{Bogomolny:1975de,Prasad:1975kr}
\begin{equation*}
    \frac{d\phi(x)}{dx} = \pm \sqrt{2V_{n,\epsilon} \big(\phi(x)\big)}.
\end{equation*}
The BPS equation is separable and can be put in the integral form
\begin{equation}
    \int_{\phi(x_0)}^{\phi(x)} \frac{d\phi}{\sqrt{2V_{n,\epsilon}(\phi)}} = \pm (x - x_0) \label{eq:BPS_integral}
\end{equation}
for some point $x_0$.
The solutions for the upper sign have positive $\mathcal{Q}$ and are called kinks and solutions for the lower sign have negative $\mathcal{Q}$ and are called antikinks.
Since the field equation has spatial reflection symmetry, for each kink profile $\phi(x)$, the corresponding antikink is given simply by $\phi(-x)$.
When there is only two global minima, the model has a single kink and antikink solutions.
However, for the triple well cases, we have a kink interpolating the vacua $-1$ and $0$ and a second one interpolating the vacua $0$ and $1$, together with their corresponding antikinks.
We label these kinks and antikinks by the pairs of vacua they interpolate, i.e., $(\phi_L, \phi_R)$ where $\phi_{L/R} \in \{-1, 0, 1\}$.

For any vacuum structure, the potential $V_{n,\epsilon}(\phi)$ has a V-shape near its minima. Therefore, the kink and antikink solutions will be contained in a compact region of space of size $L_{(\phi_L, \phi_R)}(n,\epsilon)$.
By convention, we consider the kink support to be centered at the origin, so its support is the interval $x \in [ -L_{(\phi_L, \phi_R)}(n,\epsilon)/2, L_{(\phi_L, \phi_R)}(n,\epsilon)/2]$.
From this boundary conditions, we can use equation~\eqref{eq:BPS_integral} to determine the kink support size
\begin{equation}
    L_{(\phi_L, \phi_R)}(n,\epsilon) = \int_{\phi_L}^{\phi_R} \frac{d\phi}{\sqrt{2V_{n,\epsilon}(\phi)}}
    \label{eq:kink_size}
\end{equation}
for $\phi_R > \phi_L$.
The corresponding antikink will have the same support size.
Since the kink solution $\phi(x)$ is static, we can associate the mass of the soliton with its rest energy.
Since the scalar field in our model has energy density
\begin{equation}
    \rho(t, x) = \frac{1}{2} \big(\partial_t \phi(t, x) \big)^2 + \frac{1}{2} \big(\partial_x \phi(t, x) \big)^2 + V_{n,\epsilon}\big( \phi(t, x) \big),
    \label{eq:energy-density}
\end{equation}
we can calculate the mass of a static solution as
\begin{equation*}
    M_{(\phi_L, \phi_R)}(n,\epsilon) = \int_{-L_{(\phi_L, \phi_R)}(n,\epsilon)/2}^{L_{(\phi_L, \phi_R)}(n,\epsilon)/2} dx \left[\frac{1}{2} \left(\frac{d\phi(x)}{dx}\right)^2 + V_{n,\epsilon}\big(\phi(x)\big) \right].
\end{equation*}
By using the BPS equation, this integral can be simplified to
\begin{equation}
    M_{(\phi_L, \phi_R)}(n,\epsilon) = \int_{\phi_L}^{\phi_R} d\phi \, \sqrt{2 V_{n,\epsilon}(\phi)}.
    \label{eq:kink_mass}
\end{equation}
The mass of the antikink is the same.

\subsection{\boldmath \texorpdfstring{$(-1,0)$}{(-1,0)} and \texorpdfstring{$(0,1)$}{(0,1)} kinks}

We will first discuss the case of the kinks interpolating the pairs of vacua $(-1,0)$ and $(0, 1)$.
These kink solutions are possible for the cases $\epsilon = 0$, when $\phi = 0$ is a global minimum of the potential together with $\phi = \pm 1$.
The potential has a triple well shape, similar to the $\phi^6$ model~\cite{Dorey:2011yw,Marjaneh:2017mko,Weigel:2013kwa}, and is given by the expression
\begin{equation*}
    V_{n,0}(\phi) = \frac{1}{2} \abs{\phi} \abs*{\abs{\phi}^{2n} - 1}.
\end{equation*}
For the case $\epsilon = 0$, it is enough to discuss the $(0, 1)$ kink, since the $(-1, 0)$ can be obtained as $\phi_{(-1, 0)}(x) = -\phi_{(0,1)}(-x)$ due to the symmetry of the potential under $\phi \to -\phi$.

For the $(0,1)$ sector, the BPS equation yields the following equation for the kink profile:
\begin{equation}
    2 \sqrt{\phi_{(0,1)}(x)} \, _2F_1\left(\frac{1}{2},\frac{1}{4 n};1+\frac{1}{4 n};\big( \phi_{(0,1)}(x) \big)^{2 n}\right) = x + \frac{L_{(0,1)}(n,0)}{2}
    \label{eq:BPS_triple_well}
\end{equation}
where ${}_2 F_1$ is the hyper-geometric function.
A closed form solution can be obtained for the special case $n=1/2$
\begin{equation*}
    \phi_{(0,1)}(x) =
    \begin{cases}
        0 & \text{ if } x < \mathopen{}-\frac{\pi}{2},\\
        \frac{1}{2} (1 + \sin x) & \text{ if } \mathopen{}-\frac{\pi}{2} \leq x \leq \frac{\pi}{2}, \\
        1 & \text{ if } x > \frac{\pi}{2}.
    \end{cases}
\end{equation*}
However, in the general case the kink profile has to be obtained numerically from the roots of equation~\eqref{eq:BPS_triple_well}.
We present some plots for the kink profile in figure~\ref{fig:kinks/(0,1)/profile}.

\begin{figure}
    \includegraphics{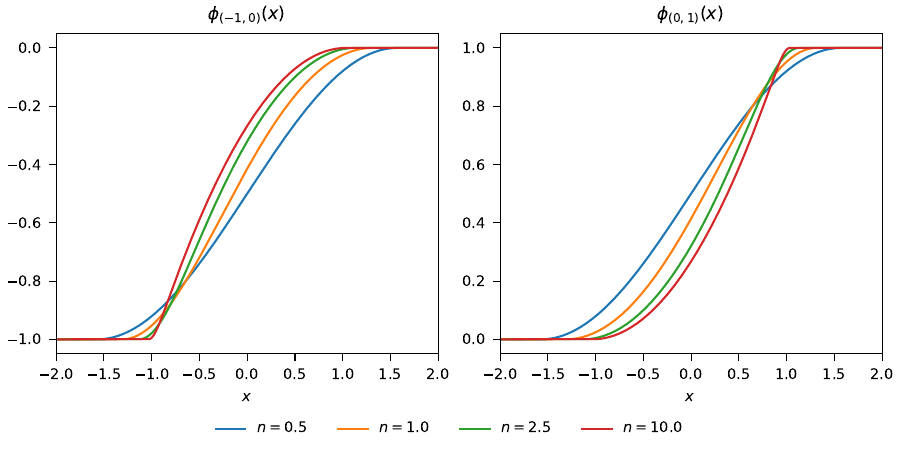}
    \caption{Kink profiles $\phi_{(-1,0)}(x)$ and $\phi_{(0,1)}(x)$ for the case $\epsilon = 0$.}
    \label{fig:kinks/(0,1)/profile}
\end{figure}

The $(0,1)$ kink has support of size
\begin{equation*}
    L_{(0,1)}(n,0) = 2 \sqrt{\pi } \frac{\Gamma \left(1+\frac{1}{4 n}\right)}{\Gamma \left(\frac{1}{2}+\frac{1}{4 n}\right)},
\end{equation*}
which is a monotonically decreasing function on $n$, with limiting values $L_{(0,1)}(1/2,0) = \pi$ and $L_{(0,1)}(\infty,0) = 2$.
The $(0,1)$ kink has mass
\begin{equation*}
    M_{(0,1)}(n,0) = \frac{\sqrt{\pi }}{3} \frac{\Gamma\left(1+\frac{3}{4n}\right)}{\Gamma \left(\frac{3}{2}+\frac{3}{4n}\right)},
\end{equation*}
which increases monotonically with $n$ and has limiting values $M_{(0, 1)}(1/2,0) = \pi/8$ and $M_{(0,1)}(\infty,0) = 2/3$.
The $(-1, 0)$ kink will have the same support size and mass.

Note that the $(-1,0)$ and $(0,1)$ kinks are only symmetric under reflection for the special case $n=1/2$.
For larger $n$, the potential reaches the vacuum $\phi=0$ differently from the $\phi=\pm 1$ vacua.
This is reflected in the kink profiles approach to each vacuum.
As $n$ increases, the kink profile approaches $\phi = \pm 1$ in an increasingly sharp way, while the approach to $\phi = 0$ has the same behavior for all $n$.

\subsection{\boldmath \texorpdfstring{$(-1,1)$}{(-1,1)} kinks}

For the $\epsilon = 0$ case, the compactness of $\phi_{(0,1)}(x)$ allows us to build a larger kink for the $(-1,1)$ sector as a superposition
\begin{equation*}
    \phi_{(-1, 1)}(x) = \phi_{(-1, 0)}(x+a) + \phi_{(0, 1)}(x-a)
\end{equation*}
provided that $a \geq L_{(0, 1)}(n,0) / 2$, i.e., that the $(-1, 0)$ and $(0, 1)$ kinks do not overlap.
In this case, we refer to the $(-1, 0)$ and $(0, 1)$ kinks as subkinks.
The size of the $(-1,1)$ kink is $L_{(0,1)}(n,0) + 2a \geq 2L_{(0,1)}(n,0)$.
By convention, we take $a = L_{(0, 1)}(n,0) / 2$, so that the subkinks have zero separation.
For larger separations, we treat each subkink as individual kinks.
Since the $(-1,0)$ and $(0,1)$ subkinks do not interact when their supports do not overlap, the mass of the $(-1, 1)$ kink is simply $2M_{(0,1)}(n,0)$.

For $\epsilon > 0$ we do not have $(-1,0)$ and $(0, 1)$ kinks, so the $(-1, 1)$ kink has to be obtained by directly solving the BPS equation with the appropriate boundary conditions.
Note that this includes both the cases when $\phi=0$ is a local minimum, or false vacuum, and the cases when $\phi = 0$ is a maximum of the potential.
In general, the integral form of the BPS equation~\eqref{eq:BPS_integral} has no closed form solution, so the kink profile has to be numerically obtained.
We present plots for the kink profile in figure~\ref{fig:kinks/(-1,1)/profile}.

\begin{figure}
    \includegraphics{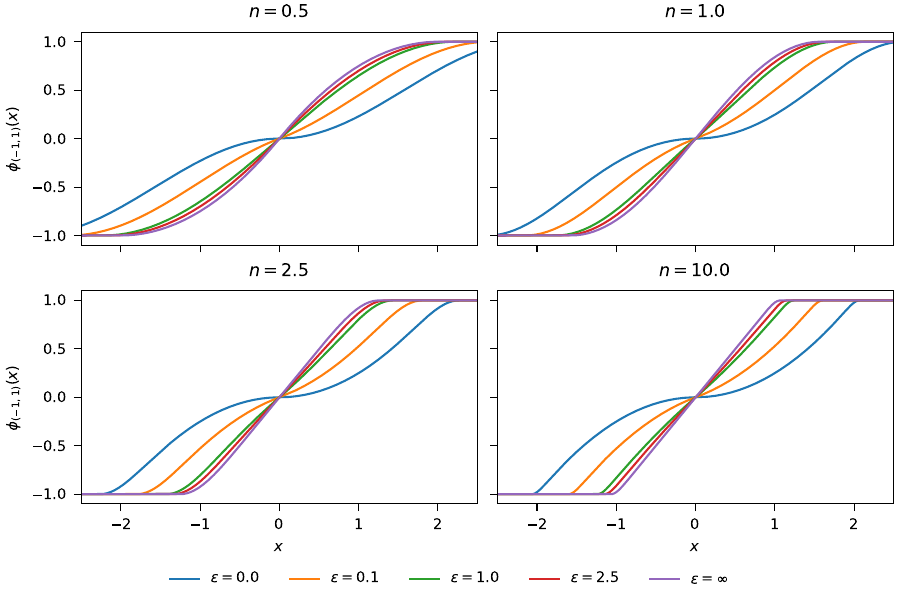}
    \caption{Kink profile $\phi_{(-1,1)}(x)$ for cases with $\epsilon > 0$.}
    \label{fig:kinks/(-1,1)/profile}
\end{figure}

Due to the presence of the false vacuum, the $(-1,1)$ kink for small $\epsilon > 0$ can be described as predominantly a superposition of two subkinks in the $(-1,0)$ and $(0,1)$ sectors.
This can be clearly seen in the energy density plots in figure~\ref{fig:kinks/(-1,1)/energy_density}.
For the cases when $\phi = 0$ is a local minimum of the potential, the energy density features two maxima, corresponding to the two subkinks.
However, as we increase $\epsilon$, these maxima converge until there is only one maximum for $\epsilon = \infty$.

\begin{figure}
    \includegraphics{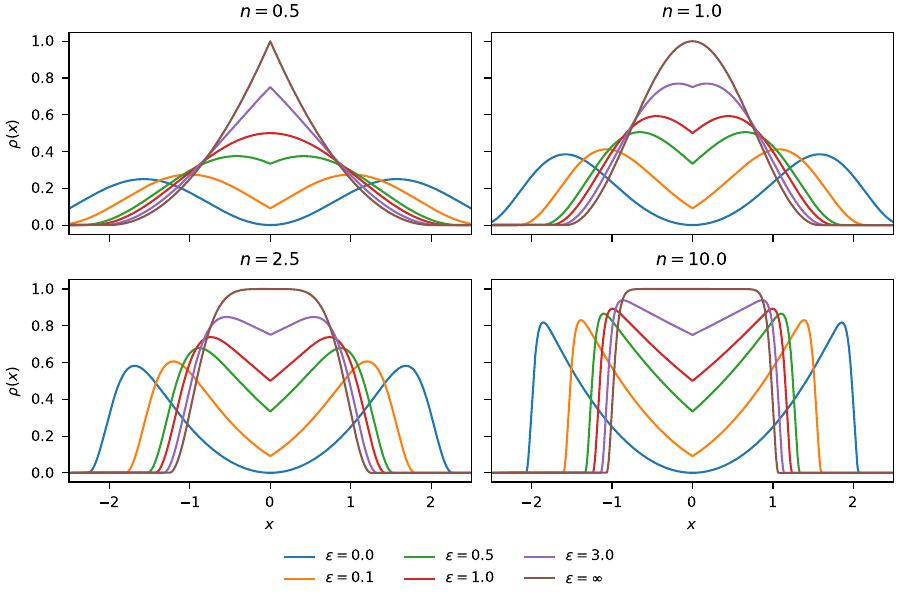}
    \caption{Energy density $\rho(x)$ for the $(-1,1)$ kink.}
    \label{fig:kinks/(-1,1)/energy_density}
\end{figure}

In the case $n = 1/2$, we are able to solve the integral in equation~\eqref{eq:BPS_integral} and analytically obtain an expression for kink profile, which is
\begin{equation*}
    \phi_{(-1, 1)}(x) =
    \begin{cases}
        -1 & \text{ if } x < -\frac{L}{2},\\
        -\sin ^2\left(\frac{x}{2 \sqrt{1+\epsilon}}-\arctan\left(\sqrt{\epsilon }\right)\right)+\epsilon  \cos ^2\left(\frac{x}{2 \sqrt{1+\epsilon}}-\arctan\left(\sqrt{\epsilon }\right)\right) &\text{ if } -\frac{L}{2} \leq x < 0,\\
        \sin ^2\left(\frac{x}{2 \sqrt{1+\epsilon}}+\arctan\left(\sqrt{\epsilon }\right)\right)-\epsilon  \cos ^2\left(\frac{x}{2 \sqrt{1+\epsilon}}+\arctan\left(\sqrt{\epsilon }\right)\right)  &\text{ if } 0 \leq x \leq \frac{L}{2},\\
        1 & \text{ if } x > \frac{L}{2}
    \end{cases}
\end{equation*}
where $L = L_{(-1,1)}(1/2,\epsilon)$ is kink support size
\begin{equation*}
    L_{(-1,1)}\left(\frac{1}{2}, \epsilon \right) = 2 \sqrt{1+\epsilon} \left[\pi -2 \arctan\left(\sqrt{\epsilon }\right)\right].
\end{equation*}
We can also find an expression for the kink mass:
\begin{equation*}
    M_{(-1,1)}(1/2,\epsilon) = \frac{(1 + \epsilon)^2 \arccot\left(\sqrt{\epsilon }\right)+(1-\epsilon) \sqrt{\epsilon }}{2 \sqrt{1 + \epsilon}}.
\end{equation*}

Another case we can analyze analytically is the limit $\epsilon \to \infty$.
In this case the potential has a double well shape and is given by the expression
\begin{equation*}
    V_{n,\infty}(\phi) = \frac{1}{2} \abs*{\abs{\phi}^{2n} - 1}.
\end{equation*}
This potential has also been studied in the context of a modified Starobinsky model~\cite{Bazeia:2024htu}.
In this case, the BPS equation for $-1 < \phi < 1$ simplifies to
\begin{equation*}
    \phi_{(-1,1)}(x) \, {}_2 F_1 \left( \frac{1}{2}, \frac{1}{2n}; 1 + \frac{1}{2n}; \abs*{\phi_{(-1,1)}(x)}^{2n} \right) = x.
\end{equation*}
There are three special cases for which the hyper-geometric function can be simplified and a closed form kink profile can be determined:
\begin{itemize}
    \item Case $n = 1/2$:
    \begin{equation*}
        \phi_{(-1,1)}(x) =
        \begin{cases}
            -1 & \text{ if } x < \mathopen{}-2, \\
            x + \frac{x^2}{4} & \text{ if } \mathopen{}-2 \leq x < 0,\\
            x - \frac{x^2}{4} & \text{ if } 0 \leq x \leq 2,\\
            1 &\text{ if } x > 2.
        \end{cases}
    \end{equation*}
    \item Case $n = 1$:
    \begin{equation*}
        \phi_{(-1,1)}(x) =
        \begin{cases}
            -1 & \text{ if } x < \mathopen{}-\frac{\pi}{2}, \\
            \sin x & \text{ if } \mathopen{}-\frac{\pi}{2} \leq x \leq \frac{\pi}{2},\\
            1 &\text{ if } x > \frac{\pi}{2}.
        \end{cases}
    \end{equation*}
    \item Case $n \to \infty$:
    \begin{equation*}
        \phi_{(-1,1)}(x) =
        \begin{cases}
            -1 & \text{ if } x < \mathopen{}-1, \\
            x & \text{ if } \mathopen{}-1 \leq x \leq 1, \\
            1 & \text{ if } x > 1. \\
        \end{cases}
    \end{equation*}
\end{itemize}
We can also calculate the kink support size as function of $n$ in the $\epsilon \to \infty$ limit and find the expression
\begin{equation*}
    L_{(-1,1)}(n,\infty) = 2\sqrt{\pi} \, \frac{\Gamma\left(1+\frac{1}{2n}\right)}{\Gamma\left(\frac{1}{2}+\frac{1}{2n}\right)}
\end{equation*}
which is a monotonically decreasing function of $n$, starting at $L_{(-1,1)}(1/2,\infty) = 4$ and going asymptotically towards $L_{(-1,1)}(\infty,\infty) = 2$.
Similarly, the kink mass can be determined from equation~\eqref{eq:kink_mass} as
\begin{equation*}
    M_{(-1,1)}(n,\infty) = \sqrt{\pi} \frac{\Gamma \left(1+\frac{1}{2 n}\right)}{\Gamma \left(\frac{3}{2}+\frac{1}{2 n}\right)} = \frac{n}{n + 1} L_{(-1,1)}(n,\infty)
\end{equation*}
which is a monotonically increasing function of $n$, starting at $M_{(-1,1)}(1/2,\infty) = 4/3$ and going asymptotically towards $M_{(-1,1)}(\infty,\infty) = 2$.

For the more general case, the kink support size and its mass were calculated numerically from equations~\eqref{eq:kink_size} and~\eqref{eq:kink_mass} and presented as color map plots in figure~\ref{fig:kinks/(-1,1)/size_and_mass}.
Both quantities depend monotonically on the parameters: the support size decreases for large $n$ or $\epsilon$, while the mass increases with $n$ and $\epsilon$.

\begin{figure}
    \includegraphics{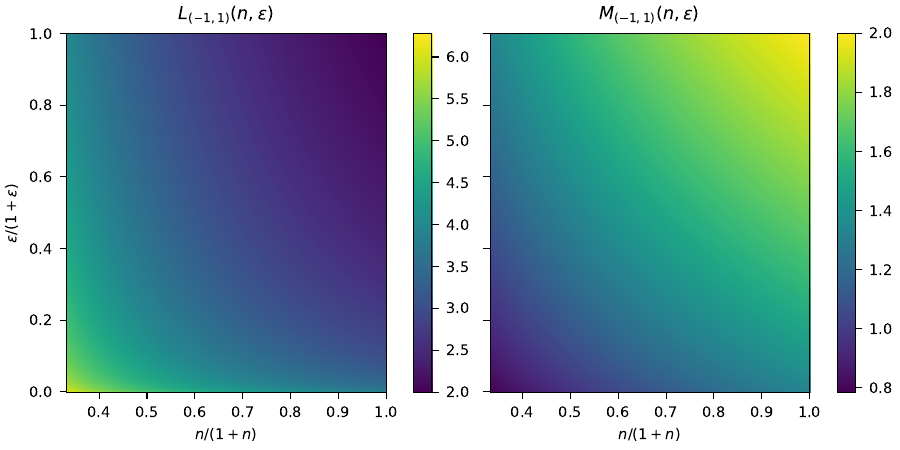}
    \caption{Support size $L_{(-1,1)}(n,\epsilon)$ and mass $M_{(-1,1)}(n,\epsilon)$ for the $(-1,1)$ kink as functions of $n / (1 + n)$ and $\epsilon / (1 + \epsilon)$. We change variables for the plot to be able to present the full dependence on the parameters, including the limits towards infinity.}
    \label{fig:kinks/(-1,1)/size_and_mass}
\end{figure}

\section{Kink internal modes}
\label{sec:internal-modes}

In this section, we start to analyze the dynamics of our compact kinks.
In particular, we look for space-time dependent solutions of the field equation by considering small oscillating perturbations on top of the previous static kink profiles, so that the field can be written as
\begin{equation}
    \phi(t, x) = \phi_{(\phi_L, \phi_R)}(x) + \chi(x) \cos \omega t,
    \label{eq:perturbed-field}
\end{equation}
where $\chi(x)$ and $\omega$ are the perturbation profile and its angular frequency, respectively.
The corresponding solutions for the perturbation profile $\chi(x)$ are known as the kink internal modes.
Following the same steps as in the standard procedure, we plug the field expression~\eqref{eq:perturbed-field} into the Euler-Lagrange equation~\eqref{eq:EL} and perform a Taylor expansion in the potential derivative.
Since the perturbation is small, we keep only the linear terms in $\chi(x)$ in the resulting expression, leading to
\begin{multline*}
    \underbrace{\partial_t^2 \phi_{(\phi_L, \phi_R)}(t, x) - \partial_x^2 \phi_{(\phi_L, \phi_R)}(t, x) + V'_{n,\epsilon}\big(\phi_{(\phi_L, \phi_R)}(t, x)\big)}_{0}\\
    + \cos(\omega t) \left[ -\omega^2 - \frac{d^2}{dx^2} + V''_{n,\epsilon}\big(\phi_{(\phi_L, \phi_R)}(x)\big) \right] \chi(x) = 0.
\end{multline*}
Hence, the internal modes and their frequencies can be obtained by solving the eigenvalue problem
\begin{equation}
    \left[- \frac{d^2}{dx^2} + U_s(x) \right] \chi(x) \label{eq:eigenproblem}
    = \omega^2 \chi(x)
\end{equation}
where the function $ U_s(x) := V''_{n,\epsilon}\big(\phi_{(\phi_L, \phi_R)}(x)\big) $ is called the stability potential.

Considering derivatives in the distributional sense, such that $\frac{d}{dz} \sgn z = 2\delta(z)$, we can compute the potential second derivative as
\begin{equation*}
    V''_{1/2,\epsilon}(\phi) = \frac{1-\epsilon}{1+\epsilon}\delta(\phi) + \delta(\abs{\phi}-1) + \frac{\sgn^2(\phi) \sgn(\abs{\phi} - 1)}{1 + \epsilon}
\end{equation*}
for the case $n = 1/2$ and
\begin{equation*}
    V''_{n,\epsilon}(\phi) = \frac{\delta(\phi)}{1+\epsilon} + 4 n^2 \, \delta\!\left(\abs{\phi} ^{2 n}-1\right) + \frac{n \sgn^2(\phi) \abs{\phi} ^{2 n-2} \sgn\left(\abs{\phi} ^{2 n}-1\right) ((2 n + 1) \abs{\phi} +\epsilon (2n-1) )}{1+\epsilon}
\end{equation*}
for $n>1/2$.
In both cases, the stability potential exhibits Dirac deltas with critical points at the vacua $\phi = \pm 1$ for all values of $\epsilon$.
Those delta functions lead to infinite barriers in the stability potential for the regions where the kink profile is equal to $\pm 1$.

In the case $n = 1/2$, for $\epsilon\neq1$, a delta distribution is also present in $V''_{1/2,\epsilon}(\phi)$ for the critical point $\phi = 0$.
For $n=1/2$ and $\epsilon < 1$, $\phi = 0$ corresponds to a minimum for the potential, and the delta brings either a sharp barrier or an infinite wall to the stability potential.
For $n=1/2$ and $\epsilon > 1$, $\phi = 0$ corresponds to a maximum, therefore the term $\delta(\phi)$ leads to a delta well in the stability potential.
In the case $n>1/2$, $V''_{n,\epsilon}(\phi)$ has a Dirac delta at the critical point $\phi = 0$ for $\epsilon < \infty$, contributing either with a delta barrier or an infinite wall to the stability potential.

Note that we performed the standard series expansion procedure in the field equation up to first order in $\chi$, even though the potential is a non-analytic function and $V''_{n,\epsilon}(\phi)$ includes delta distributions.
This is possible due to the fact that the non-analytic potential can be interpreted as the hyper-massive limit of a smooth potential parametrized by the mass of small perturbations around the vacua.
Since the expansion holds for arbitrarily very high masses, the infinite mass limit is well-defined.
A more detailed account on that issue, with further discussion, can be found in the previous works~\cite{Bazeia:2014hja,Hahne:2024qby}.

Another subtle point relates to the presence of radiation.
The infinite wall at the stability potential could lead to the impression that no radiation is possible.
However, the linearization procedure implemented here does not account for the full spectrum of excitations around the static kink solution.
Large enough perturbations are influenced by higher order terms which, in some cases, can lead to the decoupling of radiation from the topological defect.
That radiation is composed of compact oscillating packets known as oscillons.
Oscillons are solutions for small amplitude configurations in models with non-analytic potentials~\cite{Arodz:2007jh} that dominate the radiation spectrum of models with non-analytic potentials~\cite{Hahne:2019ela}.
In fact, the emission of oscillons by excited compact kinks has also been explicitly observed~\cite{Hahne:2022wyl}.

\subsection{\boldmath Modes of \texorpdfstring{$(0, 1)$}{(0,1)} kinks}

Starting with the $(0,1)$ kinks for the case $\epsilon = 0$, we calculated the stability potential numerically and presented its graph in figure~\ref{fig:kinks/(0,1)/stability_potential}.
For the general case, the stability potential can be obtained only numerically, as we do not have closed form expressions for the kink profile.
Moreover, since the kink profile depends on the potential parameter $n$, so does the stability potential.
Due to the Dirac distributions appearing in $V''(\phi)$, the stability potential displays infinite barriers at the kink borders.
This causes the internal modes to also be compact, with the same support as the static kink solution.

\begin{figure}
    \includegraphics{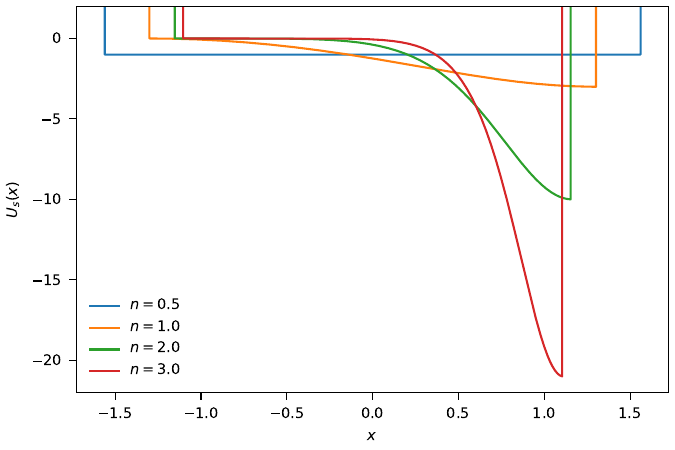}
    \caption{Stability potential $U_s(x)$ for the $(0,1)$ kinks with different values of $n$.}
    \label{fig:kinks/(0,1)/stability_potential}
\end{figure}

By numerically solving the eigenvalue problem~\eqref{eq:eigenproblem}, we can obtain the kink's internal excitation frequencies, as well as their corresponding profiles.
We present the modes frequencies and profiles in figures~\ref{fig:kinks/(0,1)/modes_frequency} and~\ref{fig:kinks/(0,1)/modes}, respectively.
The modes have been normalized in the usual way by imposing the condition $\int dx \, \big( \chi(x) \big)^2 = 1$.
All the eigenvalues $\omega^2$ are non-negative real numbers, attesting the stability of the kink solutions.
The lowest frequency, $\omega = 0$, corresponds to the translational mode $\chi_0(x)$.
In fact, it can be checked that $\chi_0(x)$ is proportional to $\phi_{(\phi_L, \phi_R)}'(x)$.

\begin{figure}
    \includegraphics{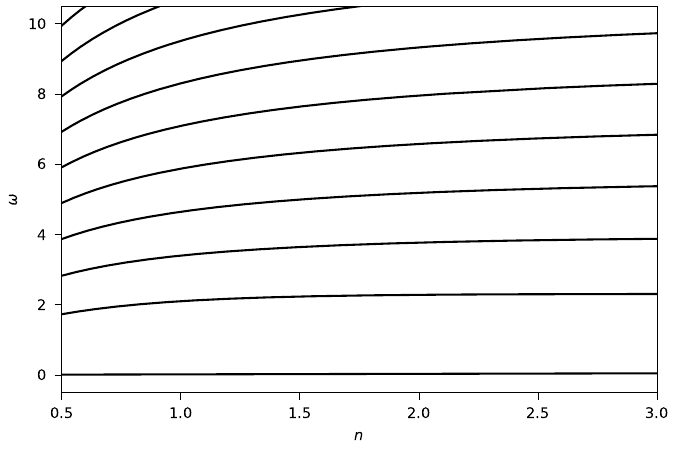}
    \caption{Frequency $\omega$ of the internal modes of the $(0,1)$ kinks as functions of $n$.}
    \label{fig:kinks/(0,1)/modes_frequency}
\end{figure}

\begin{figure}
    \includegraphics{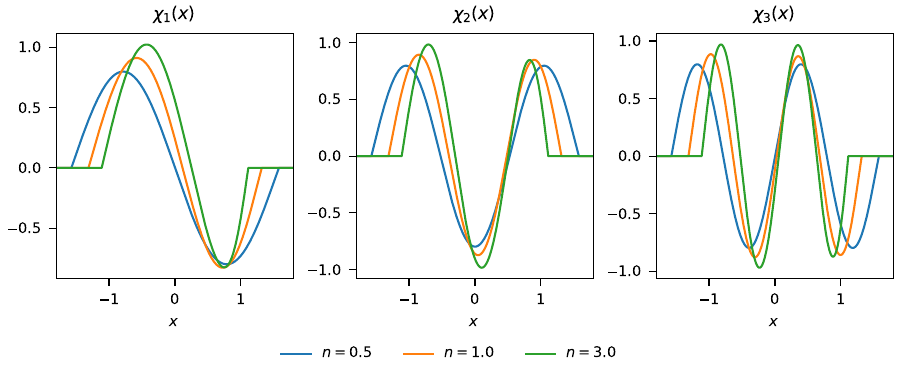}
    \caption{Profile of first three non-zeroth internal modes of the $(0,1)$ kinks for different values of $n$.}
    \label{fig:kinks/(0,1)/modes}
\end{figure}

\subsection{\boldmath Modes of \texorpdfstring{$(-1, 1)$}{(-1,1)} kinks}

For the $(-1,1)$ kinks, the Dirac deltas in $V''(\phi)$ at $\phi = \pm 1$ give rise to infinite walls in the stability potential at the ends of the kink support.
Since the kink profile has the value $\phi = 0$ at $x=0$, the stability potential can also feature a Dirac delta at that point.
In particular, for $n = 1/2$, we have a delta barrier for $\epsilon < 1$, and a delta well for $\epsilon > 1$.
For $n = 1/2$ and $\epsilon = 0$, the potential has a smooth maximum at $\phi = 0$, therefore neither barrier nor well show up at $x = 0$.
Meanwhile, for $n > 1/2$ with finite $\epsilon$, the stability potential features a delta barrier at $x = 0$.
In the limit $\epsilon\rightarrow\infty$, when $n> 1/2$, the delta at $\phi = 0$ does not contribute.
We present plots of the stability potential in figure~\ref{fig:kinks/(-1,1)/stability_potential}.

\begin{figure}
    \includegraphics{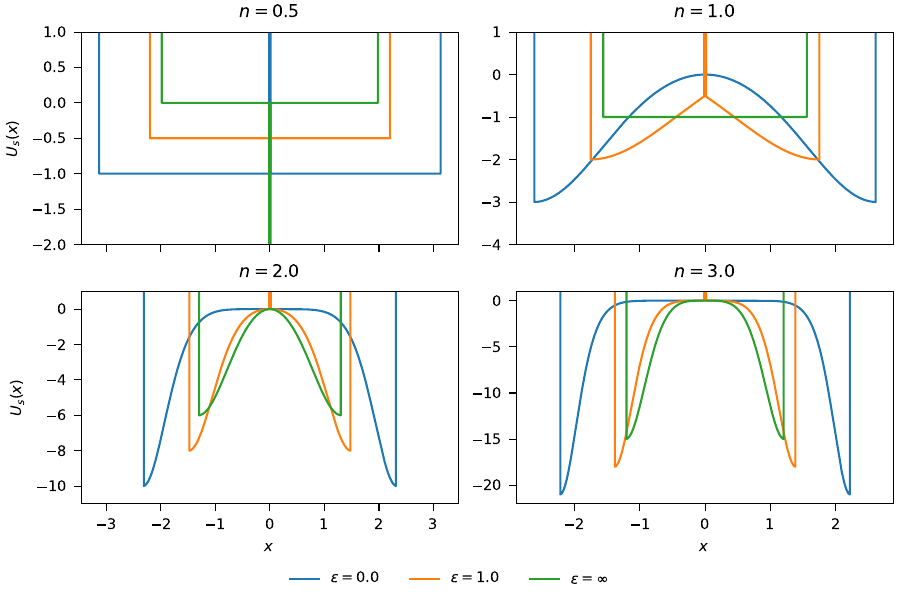}
    \caption{Stability potential $U_s(x)$ for the $(-1,1)$ kinks with different values of $n$.}
    \label{fig:kinks/(-1,1)/stability_potential}
\end{figure}

We solved the eigenfunction equation~\eqref{eq:eigenproblem} numerically to obtain the profiles and the frequencies for the modes of the $(-1, 1)$ kinks, which are presented in figures~\ref{fig:kinks/(-1,1)/modes}--\ref{fig:kinks/(-1,1)/modes_frequency_colormap}.
We observe that for $\epsilon = 0$, when the $(-1,1)$ kink is simply a superposition of the $(-1, 0)$ and $(0, 1)$ subkinks, the internal modes are also superposition of the subkinks internal modes, in either symmetric or antisymmetric fashion, as can be observed in the first row of figure~\ref{fig:kinks/(-1,1)/modes}.
This means that the frequencies' spectrum is degenerated, with $\omega_{2k} = \omega_{2k+1}$ for each $k \in \{ 0, 1, 2, \ldots \}$.
However, as we increase the value of $\epsilon$, the degeneracy is broken, with the frequencies increasing as we increase $\epsilon$, as can be seen for selected values of $n$ in figure~\ref{fig:kinks/(-1,1)/modes_frequency}.
Note that the lowest frequency remains at $\omega_0 = 0$, corresponding to the translational mode.
We also look at the dependency on the two parameters together in figure~\ref{fig:kinks/(-1,1)/modes_frequency_colormap}.
The first non-zero mode $\omega_1$ has a non-monotonic dependence on $n$, as evidenced by the level curves.

\begin{figure}
    \includegraphics{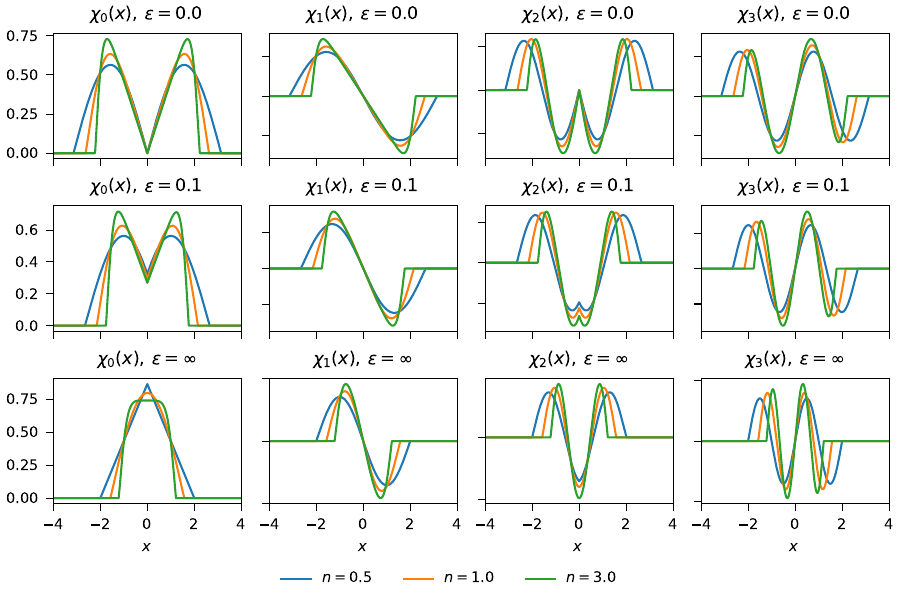}
    \caption{Profile of first four internal modes of the $(-1,1)$ kinks for different values of $n$ and $\epsilon$.}
    \label{fig:kinks/(-1,1)/modes}
\end{figure}

\begin{figure}
    \includegraphics{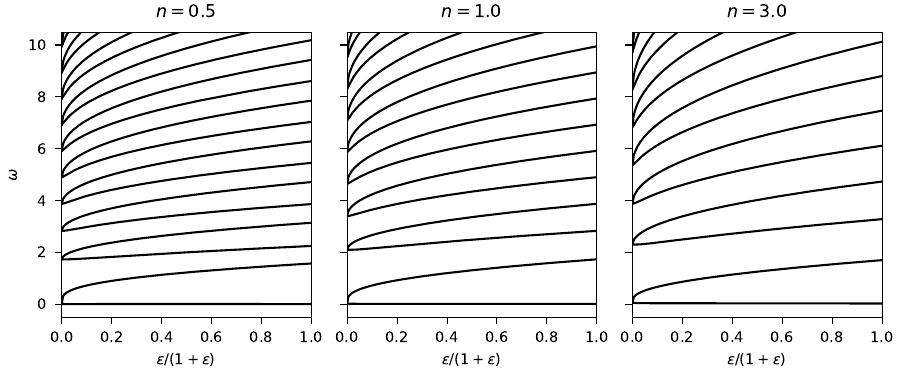}
    \caption{Frequency $\omega$ of the internal modes of the $(-1,1)$ kinks as functions of $\epsilon / (1+\epsilon)$ for some values of $n$.}
    \label{fig:kinks/(-1,1)/modes_frequency}
\end{figure}

\begin{figure}
    \includegraphics{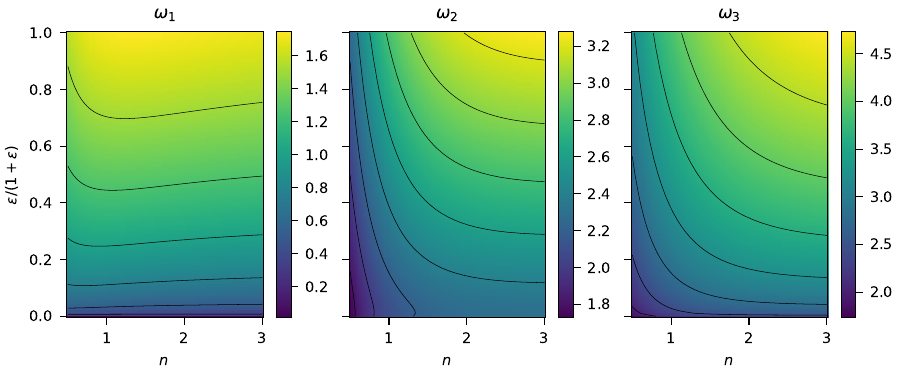}
    \caption{First three non-zero frequencies $\omega_k$ of the modes of $(-1,1)$ kinks as functions of $\epsilon / (1+\epsilon)$ and $n$. The solid lines are level curves.}
    \label{fig:kinks/(-1,1)/modes_frequency_colormap}
\end{figure}

Note that the case $\epsilon \to \infty$ and $n=1$ has closed form expressions for the modes and their frequencies, as first presented in ref.~\cite{Bazeia:2014hja}.
In that case $\omega_k^2 = k (k + 2)$, and the modes $\chi_k$ are pieces of trigonometric functions restricted to the kink support.

\section{Kink collisions}
\label{sec:collisions}

Due to the Lorentz symmetry of the model, we can construct moving kink solutions from the static solutions of section~\ref{sec:static-kinks}.
Given a static kink, with profile $\phi(x)$, the corresponding moving kink solution with velocity $v$ is simply $\phi(\gamma(x - vt))$, where $\gamma = (1-v^2)^{-1/2}$ and $\abs{v} < 1$ since we are working on natural units.
Furthermore, the compactness of the solution makes it possible to construct multi-kink configurations by simply adding the expressions for kink or antikink solutions with non-overlapping supports.
Such configurations can be taken as initial conditions to study the interaction of compact kinks during collisions.

Let us consider collisions where a kink or antikink interpolating the vacua $(\phi_L, \phi_C)$ moves from the left to the right with velocity $v$, while a kink or antikink interpolating $(\phi_C, \phi_R)$ moves from the right to the left with velocity $-v$.
We denote this configuration as a $(\phi_L, \phi_C) + (\phi_C, \phi_R)$ collision, where $\phi_{L}$, $\phi_C$, and $\phi_R$ are vacua.
Before the collision the field can be written exactly as a simple superposition of the moving soliton solutions.
When the collision begins, at $t = 0$, the field and its time derivative are given by
\begin{align}
    \phi(0, x) &= \phi_{(\phi_L, \phi_C)}(\gamma(x + x_0)) + \phi_{(\phi_C, \phi_R)}(\gamma(x - x_0)) - \phi_C, \label{eq:ic-field} \\
    \partial_t \phi(0, x) &= - v \gamma \phi'_{(\phi_L, \phi_C)}(\gamma(x + x_0)) + v \gamma \phi'_{(\phi_C, \phi_R)}(\gamma(x - x_0)), \label{eq:ic-derivative}
\end{align}
where $x_0 = L_{(\phi_L,\phi_C)}(n,\epsilon) / (2 \gamma) = L_{(\phi_C, \phi_R)}(n, \epsilon) / (2\gamma)$.
We take these as initial conditions and numerically evolve the system using the field equation~\eqref{eq:EL}.
More details about the numerical work is provided in the appendix.
We restrict ourselves to cases where both solitons have the same mass and support size, so that the initial conditions \eqref{eq:ic-field} and \eqref{eq:ic-derivative} are written in the center of momentum frame.

Due to the symmetry of the field equation under spatial reflection and under the transformation $\phi \to -\phi$, there are only four independent configurations to be considered:
\begin{enumerate}
    \item Kink-kink collisions $(-1,0) + (0, 1)$, which are equivalent to the antikink-antikink collisions $(1, 0) + (0, -1)$.
    \item Kink-antikink collisions $(0, 1) + (1, 0)$, which are equivalent to the antikink-kink collisions $(0, -1) + (-1, 0)$.
    \item Antikink-kink collisions $(1, 0) + (0, 1)$, which are equivalent to the kink-antikink collisions $(-1, 0) + (0, -1)$.
    \item Kink-antikink collisions $(-1, 1) + (1, -1)$, which are equivalent to the antikink-kink collisions $(1, -1) + (-1, 1)$.
\end{enumerate}
The first three configurations are only possible for the case of a triple well potential ($\epsilon = 0$).
Also note that for all of these configurations the initial conditions in equations~\eqref{eq:ic-field} and~\eqref{eq:ic-derivative} are either odd or even functions of the position $x$.

\subsection{\boldmath Kink-kink collisions in the \texorpdfstring{$(-1, 0) + (0, 1)$}{(-1,0) + (0,1)} sector}

First we study kink-kink collisions in the triple well case $\epsilon = 0$.
This is the only case considered with a non-zero total topological charge.
The presence of the topological charge constrains the field dynamics and allows us to predict that both solitons will survive the collision.
Furthermore, the initial conditions for this case are of the form
\begin{align*}
    \phi(0, x) &= \phi_{(-1, 0)}(\gamma(x+x_0)) + \phi_{(0,1)}(\gamma(x-x_0)) \\
    &= - \phi_{(0, 1)}(- \gamma(x+x_0)) + \phi_{(0,1)}(\gamma(x-x_0)), \\
    \partial_t \phi(0, x) &= - v \gamma \phi'_{(-1, 0)}(\gamma(x + x_0)) + v \gamma \phi'_{(0, 1)}(\gamma(x - x_0)) \\
    &= - v \gamma \phi'_{(0, 1)}(-\gamma(x + x_0)) + v \gamma \phi'_{(0, 1)}(\gamma(x - x_0)),
\end{align*}
which are odd functions of the position $x$.
Since parity is preserved in our model, we have also this dynamical constraint to the evolution of the system which implies, in particular, that $\phi(t,0) = 0$.
Therefore, the formation of additional structures in the region between the kinks is limited.

We present some simulation results as spacetime color maps for different collision velocities in figures~\ref{fig:collisions/(-1,0)+(0,1)/examples/n=0.5}--\ref{fig:collisions/(-1,0)+(0,1)/examples/n=3.0}.
We plot both the field $\phi(t, x)$ and its energy density $\rho(t,x)$ given by the expression in equation~\eqref{eq:energy-density}.
The energy density is plotted in log scale.
For the case $n=0.5$ (figure~\ref{fig:collisions/(-1,0)+(0,1)/examples/n=0.5}) the collision between the kinks is mostly elastic, with no visible emission of radiation.
This is expected, since the potential $V_{0.5,0}(\phi)$ has the same shape for $\phi \in [-1, 1]$ than the periodic potential studied in ref.~\cite{Hahne:2023dic} for $\phi \in [-2, 2]$, apart from a constant scale factor.
For that potential, the kink-kink collision is also mostly elastic, with no visible emission of radiation.

\begin{figure}[h!]
    \includegraphics{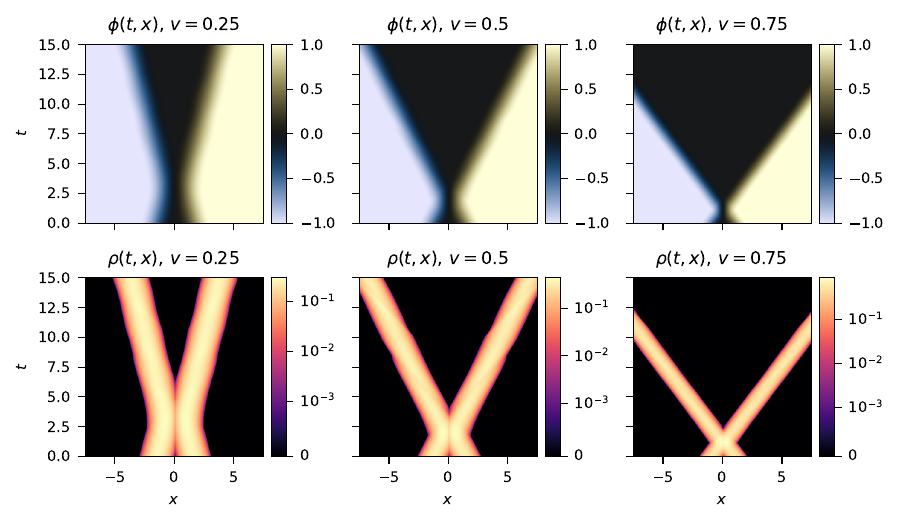}
    \caption{Field $\phi(t, x)$ and energy density $\rho(t, x)$ for $(-1,0)+(0,1)$ collisions with $\epsilon=0$, $n=0.5$, and selected values of $v$.}
    \label{fig:collisions/(-1,0)+(0,1)/examples/n=0.5}
\end{figure}

\begin{figure}[h!]
    \includegraphics{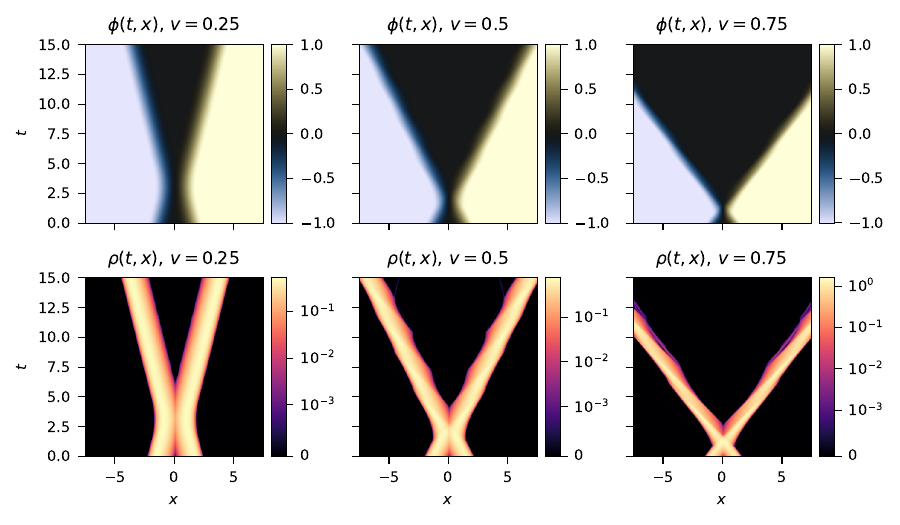}
    \caption{Field $\phi(t, x)$ and energy density $\rho(t, x)$ for $(-1,0)+(0,1)$ collisions with $\epsilon=0$, $n=1.5$, and selected values of $v$.}
    \label{fig:collisions/(-1,0)+(0,1)/examples/n=1.5}
\end{figure}

\begin{figure}[h!]
    \includegraphics{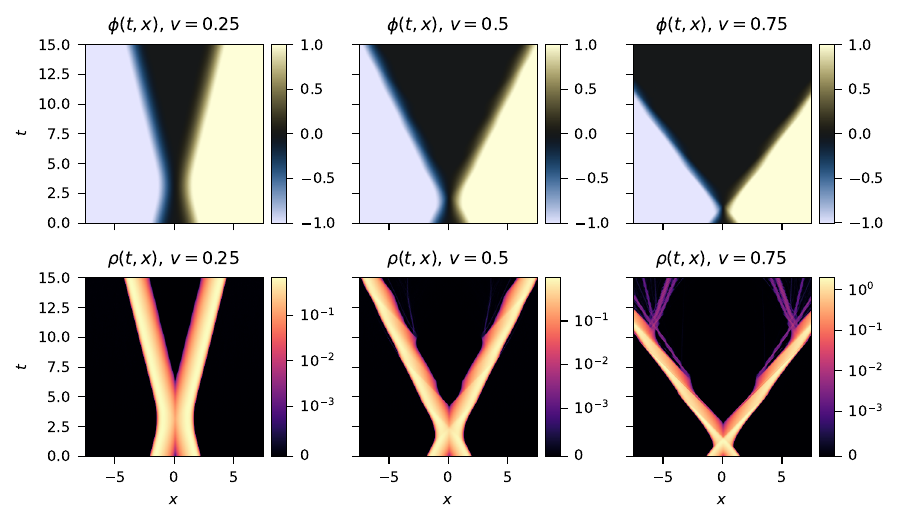}
    \caption{Field $\phi(t, x)$ and energy density $\rho(t, x)$ for $(-1,0)+(0,1)$ collisions with $\epsilon=0$, $n=3$, and selected values of $v$.}
    \label{fig:collisions/(-1,0)+(0,1)/examples/n=3.0}
\end{figure}

However, as we increase the parameter $n$, the interaction becomes more complex.
For $n=1.5$ (figure~\ref{fig:collisions/(-1,0)+(0,1)/examples/n=1.5}) the outgoing kinks are excited, as can be easily seen in the energy density plots as a vibration of the kink support.
For $n = 3$ (figure~\ref{fig:collisions/(-1,0)+(0,1)/examples/n=3.0}), the kinks emit some of their extra energy as radiation.
Curiously, the vibration of the support and the emission of radiation is visible mostly in the inner region, in which the kinks touch the $\phi = 0$ vacuum.
However, the parameter $n$ changes the approach of potential and kink profiles near the $\phi=\pm 1$ vacua.
A possible explanation is that, around $\phi = 0$, the kink is less likely to trap perturbations inside its bulk.
In models with non-analytic potentials, small perturbations on the kink bulk encounter wall like potential barriers at the kink support borders which cause radiation to be at least partially trapped inside the kink bulk~\cite{Hahne:2022wyl}.
For $n > 1/2$ the potential is more steep around the vacua $\phi = \pm 1$ than around $\phi = 0$, making the barrier effect more pronounced around $\phi = \pm 1$.
Therefore, even if increasing $n$ changes the potential mostly around $\phi = \pm 1$, the resulting radiation is more likely to escape around $\phi = 0$.

\subsection{\boldmath Kink-antikink collisions in the \texorpdfstring{$(0, 1) + (1, 0)$}{(0,1) + (1,0)} sector}

For the case of a kink-antikink collision in the $(0, 1) + (1, 0)$ sector, the initial conditions are given by
\begin{align*}
    \phi(0, x) &= \phi_{(0, 1)}(\gamma(x + x_0)) + \phi_{(1, 0)}(\gamma(x - x_0)) - 1 \\
    &= \phi_{(0, 1)}(\gamma(x + x_0)) + \phi_{(0, 1)}(\gamma(- x + x_0)) - 1,\\
    \partial_t \phi(0, x) &= - v \gamma \phi'_{(0, 1)}(\gamma(x + x_0)) + v \gamma \phi'_{(1, 0)}(\gamma(x - x_0))\\
    &= - v \gamma \phi'_{(0, 1)}(\gamma(x + x_0)) - v \gamma \phi'_{(0, 1)}(\gamma( -x + x_0)),
\end{align*}
which are even functions of position, i.e., the configuration is symmetric under spatial reflection.

We present some simulation results for different values of $n$ and $v$ in figures~\ref{fig:collisions/(0,1)+(1,0)/examples/n=0.5}--\ref{fig:collisions/(0,1)+(1,0)/examples/n=3.0}.
As usual in the case of kink-antikink collisions in non-integrable models, the collisions leads to either a capture of the soliton pair into a bion configuration, or an escape of the pair for large enough collision velocities $v$.
In all cases, there is emission of radiation, as can be seen in the energy density plots.
The radiation is emitted both to the outer and inner regions, allowing for the formation of central structures, such as shockwaves, usually short-lived, and oscillons, long-lived.

\begin{figure}
    \includegraphics{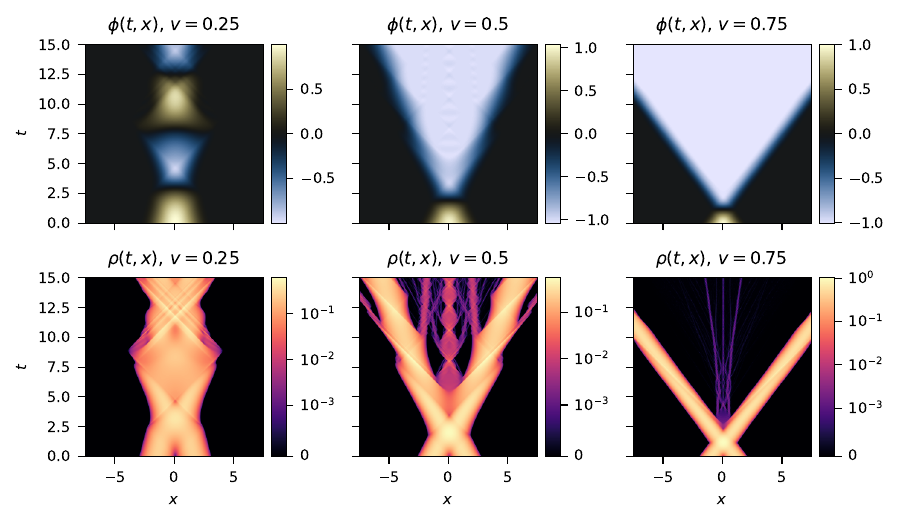}
    \caption{Field $\phi(t, x)$ and energy density $\rho(t, x)$ for $(0,1) + (1, 0)$ collisions with $\epsilon=0$, $n=0.5$, and selected values of $v$.}
    \label{fig:collisions/(0,1)+(1,0)/examples/n=0.5}
\end{figure}

\begin{figure}
    \includegraphics{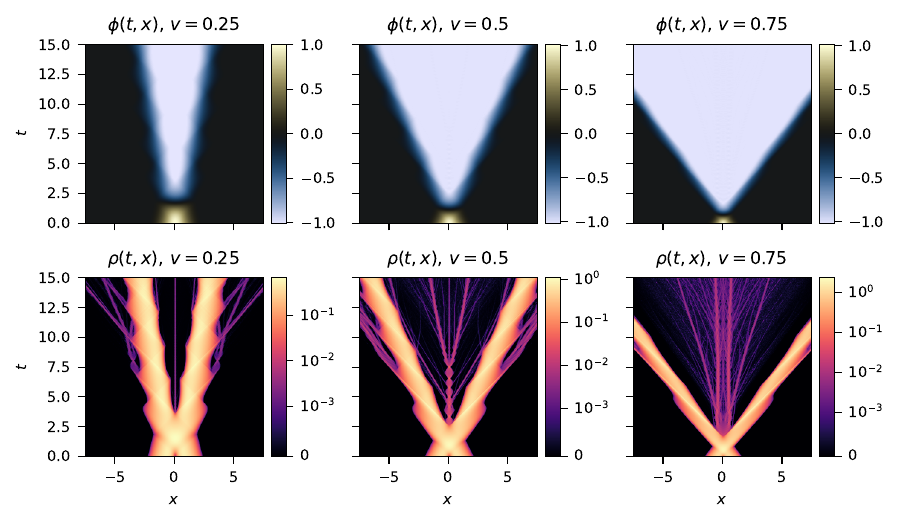}
    \caption{Field $\phi(t, x)$ and energy density $\rho(t, x)$ for $(0,1) + (1, 0)$ collisions with $\epsilon=0$, $n=1.5$, and selected values of $v$.}
    \label{fig:collisions/(0,1)+(1,0)/examples/n=1.5}
\end{figure}

\begin{figure}
    \includegraphics{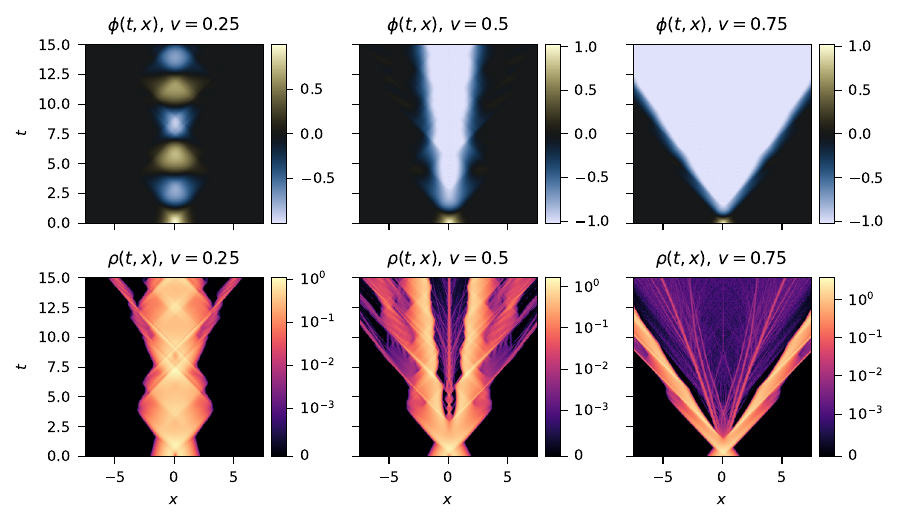}
    \caption{Field $\phi(t, x)$ and energy density $\rho(t, x)$ for $(0,1) + (1, 0)$ collisions with $\epsilon=0$, $n=3$, and selected values of $v$.}
    \label{fig:collisions/(0,1)+(1,0)/examples/n=3.0}
\end{figure}

The presence of these structures is an important characteristic of dynamical processes in non-analytic field theories.
Oscillons have been known to dominate the radiation spectrum of such models~\cite{Klimas:2018woi,Hahne:2019ela}, with decaying shockwaves being one of the main processes leading to oscillon creation~\cite{Hahne:2019odw}.
In fact, if we write $\phi = \phi_v + \delta\phi$, where $\phi_v \in \{ -1, 0, 1\}$, the field equation near the vacuum can be approximated by the signum-Gordon equation
\begin{equation*}
    \partial_t^2 \delta\phi - \partial_x^2 \delta\phi + \lambda \sgn\delta\phi = 0,
\end{equation*}
with $\lambda = 1$ for $\phi_v = 0$, and $\lambda = n$ for $\phi_v = \pm 1$.
The signum-Gordon equation has both oscillons and shockwaves as known exact solutions~\cite{Arodz:2007jh,Arodz:2005bc}.
We can introduce new coordinates $x = \lambda^{-1/2} x'$, $t = \lambda^{-1/2} t'$ to eliminate the $\lambda$ factor from the signum-Gordon equation.
Therefore, $\lambda^{-1/2}$ defines a characteristic size for the largest oscillons.
In the inner region between the outgoing solitons, the field has values close to the vacuum $\phi_v = -1$, therefore the characteristic size for the largest oscillon is $n^{-1/2}$, i.e., it decreases as $n$ increases.
This can be observed by visual comparison of figures~\ref{fig:collisions/(0,1)+(1,0)/examples/n=0.5}--\ref{fig:collisions/(0,1)+(1,0)/examples/n=3.0}, specially for the central panel with $v=0.5$.
However, since the signum-Gordon equation has scale symmetry, the scale factor will only affect the largest oscillon before higher order effects come into play.
This means that smaller oscillons appear in all cases.

Beyond the effects on radiation, the $n$ parameter affects the overall outcome of the collision.
For example, the collision with velocity $v=0.25$ results in a capture of the kink-antikink pair for $n=0.5$ and $n=3$, but it leads to an escape for the intermediary value $n=1.5$.
Furthermore, the dependence on the velocity is not usually straightforward in soliton collisions, with some models presenting an intricate fractal dependence on $v$.

We can systematically classify the outcome of a collision as either capture or escape simply by looking at the field at the central point $x=0$.
For the escape cases, $\phi(t, 0)$ will stay close to a vacuum as the solitons escape.
However, for the capture cases $\phi(t, 0)$ will oscillate between around the $\phi = 0$ vacuum due to the formation of a bion.
In figure~\ref{fig:collisions/(0,1)+(1,0)/middle_examples} we present the dependence of the field at $x=0$ as function of $t$ and $v$ for different values of the parameter $n$.
For the case $n = 1/2$, we once again reproduce the results from the periodic potential from ref.~\cite{Hahne:2023dic}.
The capture and escape cases are separated by some critical velocity.

\begin{figure}
    \includegraphics{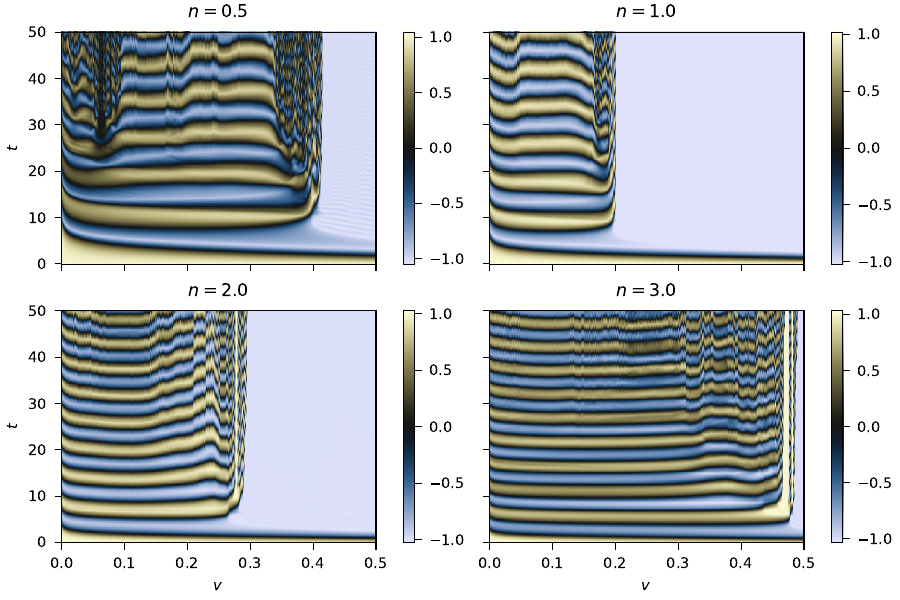}
    \caption{Field $\phi(t, 0)$ at the middle point $x=0$ between the kink-antikink pair as function of time $t$ and collision velocity $v$ for the $(0,1) + (1,0)$ sector for different values of $n$.}
    \label{fig:collisions/(0,1)+(1,0)/middle_examples}
\end{figure}

For small values of $n$, there are no cases of multi-bounces of the field followed by escape.
However, as we increase $n$, small windows of one bounce followed by escape appear slightly before the critical velocity.
We can see this as a thin yellow band in the $n=2$ and $n=3$ cases of figure~\ref{fig:collisions/(0,1)+(1,0)/middle_examples}.
This region of one bounce followed by escape differs from the other escape cases, because the vacuum in the inner region is $\phi = 1$, instead of $\phi = -1$, as the escape cases with high velocity.
We can interpret this as the two solitons approaching and receding after the collision, while the high velocity cases, for which the inner region approaches the vacuum $\phi = -1$, can be interpreted as the kinks passing through each other.

The presence of the one-bounce window is also interesting because it was predicted in the case of the periodic potential (which is similar to the present $n=1/2$ case) by a collective coordinates model based on the kink internal modes, even though it was not present at the simulations for that case.
Its presence for higher values of $n$ suggests that the low $n$ cases have some mechanism in addition to the internal modes that forbids the formation of multi-bounce windows.

\begin{figure}
    \includegraphics{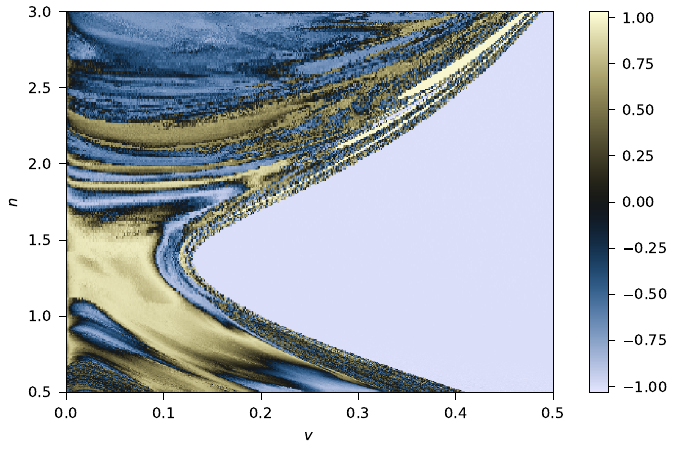}
    \caption{Field $\phi(50, 0)$ at the middle point between the kink-antikink pair as function of the collision velocity $v$ and $n$ for the $(0,1) + (1,0)$ sector.}
    \label{fig:collisions/(0,1)+(1,0)/middle_final}
\end{figure}

Another interesting feature is the non-monotonic dependence of the critical velocity on $n$.
The critical velocity has a minimum for $n = 1.36$, for which $v_c = 0.13$, as can be seen in figure~\ref{fig:collisions/(0,1)+(1,0)/middle_final}, which plots the field in the central point at a late time $t = 50$.
Note that the capture cases never completely disappear.

\subsection{\boldmath Antikink-kink collisions in the \texorpdfstring{$(1, 0) + (0, 1)$}{(1,0) + (0,1)} sector}

We now study the kink-antikink collisions in the $(1, 0) + (0, 1)$ sector.
The initial conditions are given by
\begin{align*}
    \phi(0, x) &= \phi_{(1, 0)}(\gamma(x + x_0)) + \phi_{(0, 1)}(\gamma(x - x_0)) \\
    &= \phi_{(0, 1)}(\gamma( - x - x_0)) + \phi_{(0, 1)}(\gamma(x - x_0)),\\
    \partial_t \phi(0, x) &= - v \gamma \phi'_{(1, 0)}(\gamma(x + x_0)) + v \gamma \phi'_{(0, 1)}(\gamma(x - x_0))\\
    &= v \gamma \phi'_{(0, 1)}(\gamma(-x - x_0)) + v \gamma \phi'_{(0, 1)}(\gamma(x - x_0)).
\end{align*}
Once again, the configuration is symmetric under spatial reflection.
Despite being a qualitatively similar configuration to the $(0, 1) + (1, 0)$ sector, the $(1, 0) + (0, 1)$ collision pattern is not equivalent to the former.
In particular, the boundary conditions far from the solitons are fundamentally different, approaching the outer minimum of the potential ($\phi = 1$) instead of the inner minimum ($\phi = 0$).
The distinction between the inner and outer minima of the potential is known to produce differences in kink collisions in other models, such as the $\phi^6$~\cite{Dorey:2011yw} and $\phi^8$ ones~\cite{Bazeia:2023qpf}.

\begin{figure}[h!]
    \includegraphics{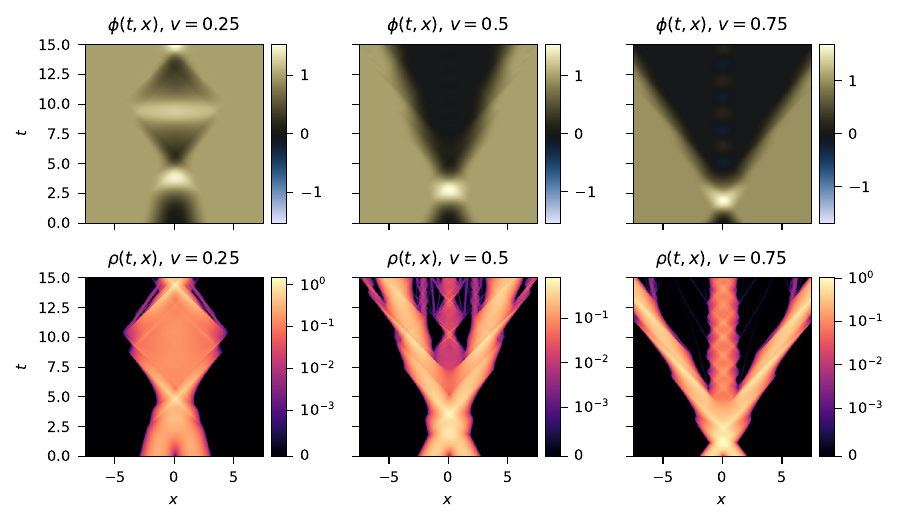}
    \caption{Field $\phi(t, x)$ and energy density $\rho(t, x)$ for $(1,0) + (0, 1)$ collisions with $\epsilon=0$, $n=0.5$, and some values of $v$.}
    \label{fig:collisions/(1,0)+(0,1)/examples/n=0.5}
\end{figure}

\begin{figure}[h!]
    \includegraphics{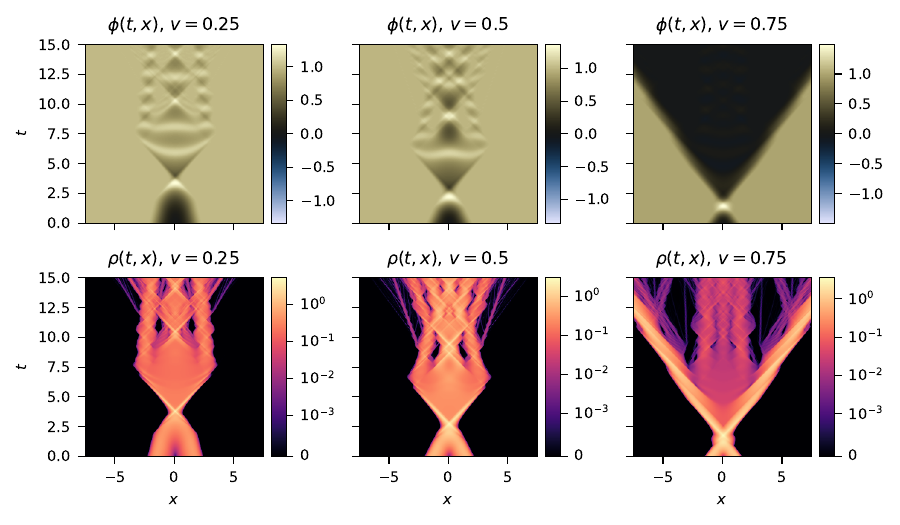}
    \caption{Field $\phi(t, x)$ and energy density $\rho(t, x)$ for $(1,0) + (0, 1)$ collisions with $\epsilon=0$, $n=1.5$, and some values of $v$.}
    \label{fig:collisions/(1,0)+(0,1)/examples/n=1.5}
\end{figure}

\begin{figure}[h!]
    \includegraphics{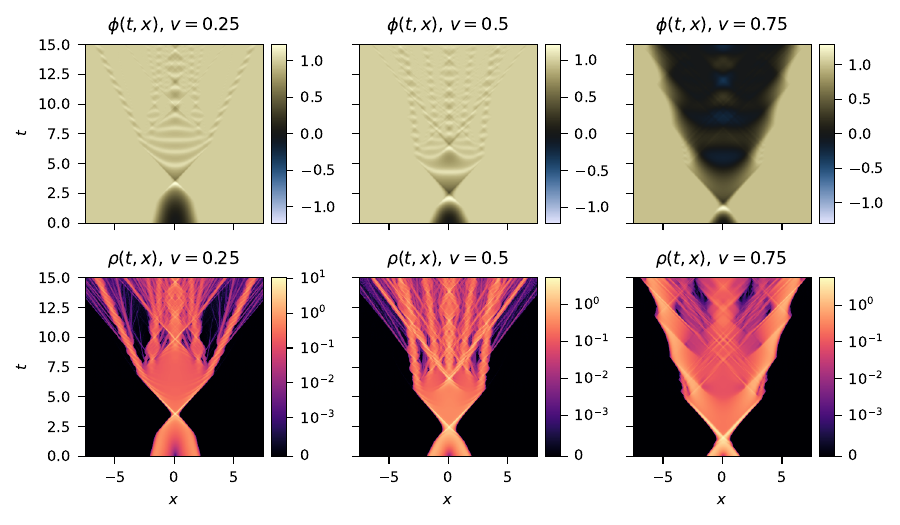}
    \caption{Field $\phi(t, x)$ and energy density $\rho(t, x)$ for $(1,0) + (0, 1)$ collisions with $\epsilon=0$, $n=3$, and some values of $v$.}
    \label{fig:collisions/(1,0)+(0,1)/examples/n=3.0}
\end{figure}

We present the simulation results for some values of $v$ and $n$ in figures~\ref{fig:collisions/(1,0)+(0,1)/examples/n=0.5}--\ref{fig:collisions/(1,0)+(0,1)/examples/n=3.0}.
Once again, the results can be broadly classified into escape or capture of the kink-antikink pair.
However, for the escape cases, the field in the region between the solitons is equal to the vacuum $\phi = 0$, the same from before the collision.
If the full process were to be described as a superposition of the original kink-antikink pair in the form
\begin{equation*}
    \phi(t, x) = \phi_{(1, 0)}(\gamma(x + a(t))) + \phi_{(0, 1)}(\gamma(x - a(t)))
\end{equation*}
where $a(t)$ is the $(0,1)$ kink position, we would have $a(0) = x_0$ and $a(t)$ would decrease during the initial moments of the collision, only to later increase again.
This means that the solitons bounced against each other, instead of passing through as it happened in the $(0,1) + (1,0)$ sector.
Another interesting finding is that the formation of central oscillons for high velocities depends on the potential parameter $n$.
While a well-defined central oscillon can be seen for $v=0.75$ in the $n=1$ case (figure~\ref{fig:collisions/(1,0)+(0,1)/examples/n=0.5}, last column), as $n$ increases the oscillon becomes less clear, while additional radiation starts to dominate (last column of figures~\ref{fig:collisions/(1,0)+(0,1)/examples/n=1.5} and \ref{fig:collisions/(1,0)+(0,1)/examples/n=3.0}).

For the capture cases, we observe the formation of large central oscillons only for small values of the parameter $n$.
When $n$ increases, the capture cases result in a rather fast decay of the whole configuration into radiation, as can be seen in the first two columns of figure~\ref{fig:collisions/(1,0)+(0,1)/examples/n=3.0}.
This effect can also be attributed to the slope of the potential around $\phi = \pm 1$, which favors the formation of small oscillons around these vacua.

\begin{figure}[h!]
    \includegraphics{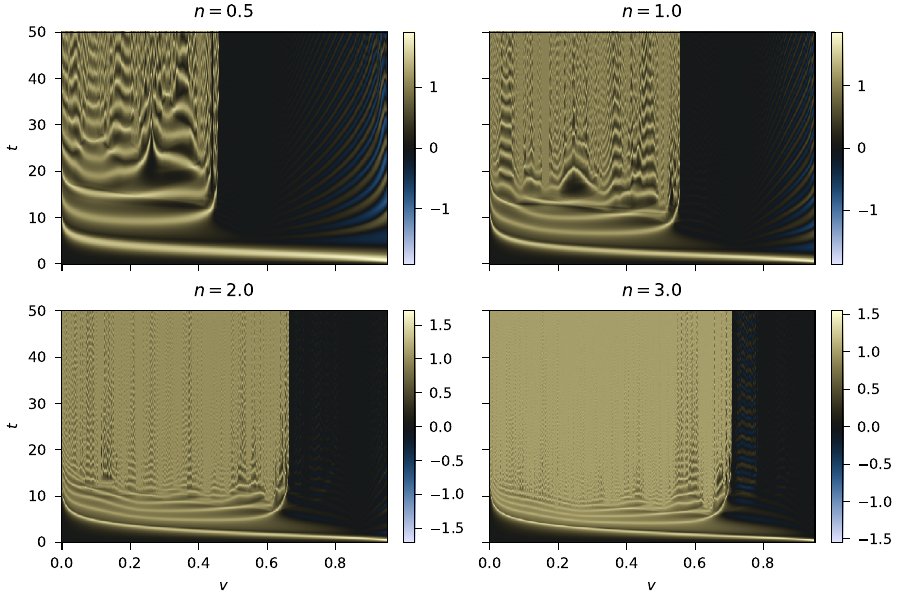}
    \caption{Field $\phi(t, 0)$ at the middle point between the antikink-kink pair as function of time $t$ and collision velocity $v$ for the $(1,0) + (0,1)$ sector with different value of $n$.}
    \label{fig:collisions/(1,0)+(0,1)/middle_examples}
\end{figure}

We take a more systematic view by looking at the dependence of $\phi(t, 0)$ on $v$ in figure~\ref{fig:collisions/(1,0)+(0,1)/middle_final}.
For $n=0.5$ (first panel), we reproduce the results from the non-analytic double well model of ref.~\cite{Hahne:2024qby}.
Note the presence of a small velocity band before the critical value for which the field bounces once before the kink-antikink pair escapes.
The formation of central oscillons is also visible for high velocities.
As we increase the value of $n$, the central oscillon for high $v$ fades out, for example in the cases $n=1.0$ and $n=2.0$.
For $n=3.0$, a central oscillon appears in the escape cases, but only for velocities slightly greater than the central velocity (first panel of the second row).
The full scans also confirm that the period of the bion formed in the capture cases decreases as $n$ increases.

\begin{figure}[h!]
    \includegraphics{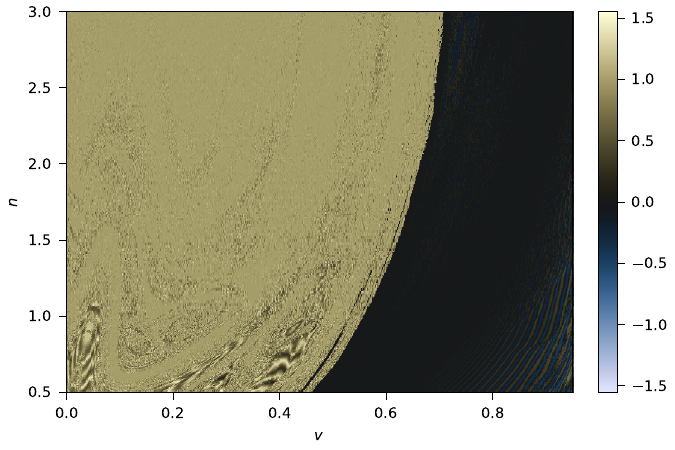}
    \caption{Field $\phi(50, 0)$ at the middle point between the antikink-kink pair as function of the collision velocity $v$ and $n$ for the $(1,0) + (0,1)$ sector.}
    \label{fig:collisions/(1,0)+(0,1)/middle_final}
\end{figure}

The effects of the parameter $n$ can also be analyzed by looking at the field at $x=0$ for a late time.
In figure~\ref{fig:collisions/(1,0)+(0,1)/middle_final}, we present the field at $t=50$ as a function of $n$ and $v$.
This scan shows that the critical velocity grows mostly monotonically with $n$ in the range $n \in  [0.5, 3]$, although with some noise.
Furthermore, the band of escape cases before the critical velocity remains visible for $n \lesssim 1.5$.

\subsection{\boldmath Kink-antikink collisions in the \texorpdfstring{$(-1, 1) + (1, -1)$}{(-1,1) + (1,-1)} sector}

For the last case, concerning the $(-1, 1) + (1, -1)$ kink-antikink collisions, the initial conditions are
\begin{align*}
    \phi(0, x) &= \phi_{(-1, 1)}(\gamma(x + x_0)) + \phi_{(1, -1)}(\gamma(x - x_0)) - 1 \\
    &= \phi_{(-1, 1)}(\gamma(x + x_0)) - \phi_{(-1, 1)}(\gamma(x - x_0)) - 1,\\
    \partial_t \phi(0, x) &= - v \gamma \phi'_{(-1, 1)}(\gamma(x + x_0)) + v \gamma \phi'_{(1, -1)}(\gamma(x - x_0))\\
    &= - v \gamma \phi'_{(-1, 1)}(\gamma(x + x_0)) - v \gamma \phi'_{(-1, 1)}(\gamma( -x + x_0))
\,.
\end{align*}

We first look at the case $\epsilon = 0$, for which the potential has a triple well shape.
For this case, the kinks $(-1, 1)$ are simply a superposition of a $(-1, 0)$ and $(0, 1)$ kinks with zero separation and no overlap.
Since each subkink is compact, they do not interact and can therefore be separated simply by acquiring different velocities.
Hence, a $(-1, 1) + (1, -1)$ collision for $\epsilon = 0$ is actually a four soliton collision $(-1, 0) + (0, 1) + (1, 0) + (0, -1)$.

\begin{figure}
    \includegraphics{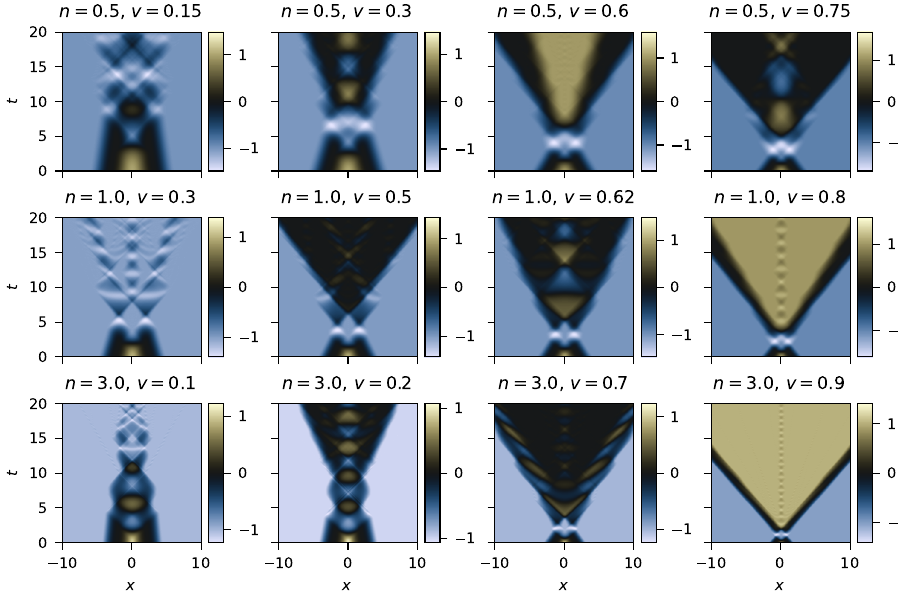}
    \caption{Field $\phi(t, x)$ for $(-1, 1) + (1, -1)$ collisions with $\epsilon = 0$ and selected values of $n$ and $v$.}
    \label{fig:collisions/(-1,1)+(1,-1)/examples/epsilon=0.0}
\end{figure}

We present some examples of the simulation results for $\epsilon = 0$ in figure~\ref{fig:collisions/(-1,1)+(1,-1)/examples/epsilon=0.0}.
The presence of four solitons generates a highly non-trivial dynamics, with several possible outcomes.
For small velocities, we observe cases with a complete annihilation of the kink-antikink pair, such as the cases $n = 0.5$ with $v = 0.15$, $n = 1$ with $v = 0.3$, and $n = 3.0$ with $v = 0.1$, displayed in the first column of the plots.
There are also cases in which only the inner subkinks $(0, 1)$ and $(1, 0)$ annihilate, but the outer subkinks $(-1, 0)$ and $(0, -1)$ are able to escape.
These include cases with a formation of a large central oscillon ($n = 0.5$ with $v = 0.75$, $n = 1$ with $v = 0.62$, and $n = 3$ with $v = 0.2$, for example), as well as cases without it ($n = 1$ with $v = 0.5$ and $n = 3$ with $v = 0.7$, for example).
The case $n = 3$ with $v = 0.7$ is specially interesting, since it features two large oscillons that accompany the outgoing solitons.
We also observe cases in which all four solitons escape, such as $n = 0.5$ with $v = 0.6$, $n = 1$ with $v = 0.8$, and $n = 3$ with $v = 0.9$, for example.
In the first two of these examples, the $(-1, 1)$ kink and the $(1, -1)$ antikink are broken into their composing subkinks, i.e., each one of the subkinks and sub-antikinks escapes with a different velocity, forming two growing regions where $\phi = 0$ (dark regions in the plots).
In the last example, $n = 3$ with $v = 0.9$, the $(-1, 1)$ kink and $(1, -1)$ antikink remain intact.

\begin{figure}
    \includegraphics{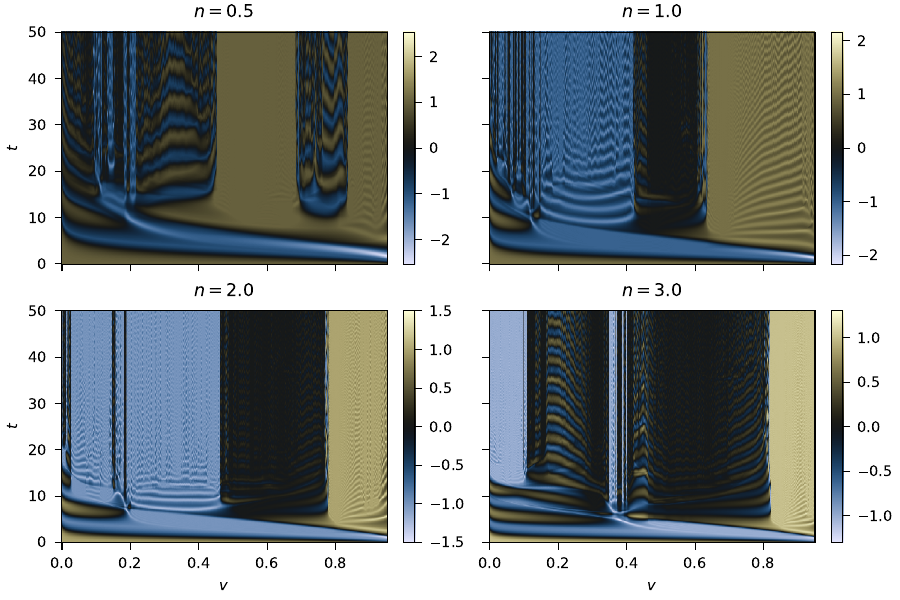}
    \caption{Field $\phi(t, 0)$ for $(-1, 1) + (1, -1)$ collisions as a function of the velocity $v$ for selected values of $n$ in the case $\epsilon = 0$, for which the potential has a triple well shape.}
    \label{fig:collisions/(-1,1)+(1,-1)/middle_examples/epsilon=0.0}
\end{figure}

We notice that the outcome depends on the velocity in a very non-trivial manner.
Even for high velocities, such as the case $n = 0.5$, $v = 0.75$, we observe partial captures, even though smaller velocities lead to full subkink escapes.
We explore this dependency in figure~\ref{fig:collisions/(-1,1)+(1,-1)/middle_examples/epsilon=0.0}, where we plot the field at the origin $x=0$ as a function of time $t$ and velocity $v$ for $\epsilon = 0$ and selected values of $n$.
In the case $n = 0.5$, there is an initial window of velocities for which there are partial or complete captures, followed by escape instances.
However,  we also observe a second window with partial captures.
For the cases $n = 1$ and $n = 2$, complete capture, partial capture, and escape events appear mostly contained in successive windows of velocity, with some smaller outlier windows within.
For these values of $n$, the capture windows with formation of large oscillons are narrow.
However, for $n=3$, there are wider windows of large oscillon formation.
The alternating of the windows is also significantly more intricate in this case.

\begin{figure}
    \includegraphics{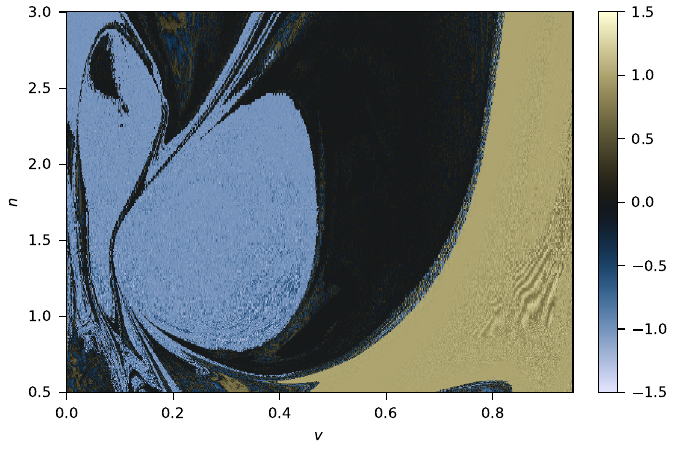}F
    \caption{Field $\phi(50, 0)$ for $(-1, 1) + (1, -1)$ collisions as function of velocity $v$ and $n$ in the case $\epsilon = 0$, for which the potential has a triple well shape.}
    \label{fig:collisions/(-1,1)+(1,-1)/middle_final/epsilon=0.0}
\end{figure}

The dependence on $n$ can be more systematically explored by looking at the field at the origin on later times.  Choosing $t = 50$, in figure~\ref{fig:collisions/(-1,1)+(1,-1)/middle_final/epsilon=0.0}, we present the values of $\phi(50, 0)$ as function of $n$ and $v$.
Looking at the lower region of the plot, we see that the formation of a second capture window is a feature mostly of values of $n$ only slightly larger than $0.5$.
In general, we see that the capture window grows with $n$, specially for partial captures.

\begin{figure}
    \includegraphics{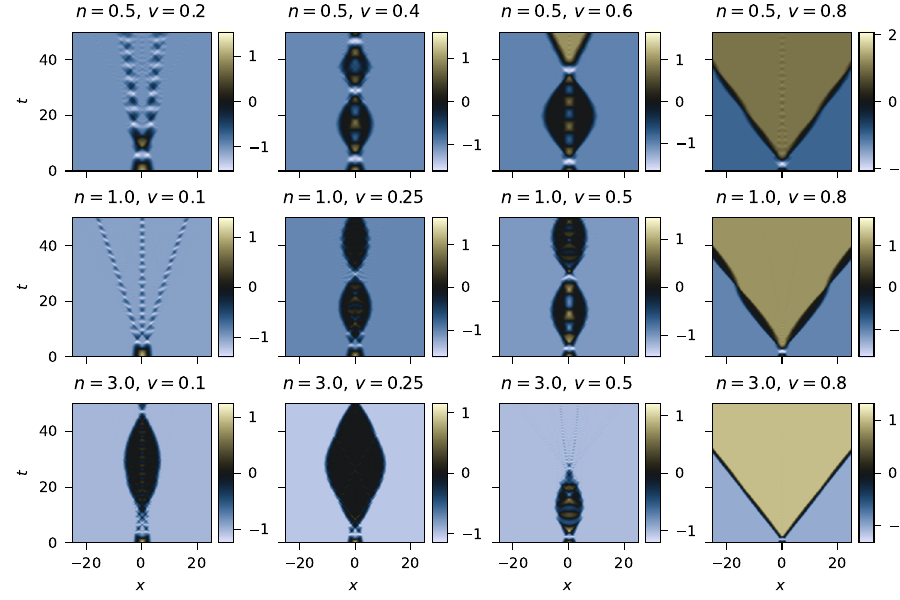}
    \caption{Field $\phi(t, x)$ for $(-1, 1) + (1, -1)$ collisions with $\epsilon = 0.1$ and selected values of $n$ and $v$.}
    \label{fig:collisions/(-1,1)+(1,-1)/examples/epsilon=0.1}
\end{figure}

Proceeding further, let us look at the collisions in the case $\epsilon = 0.1$, for which the potential features a false vacuum.
Some examples of simulations for this case are presented in figure~\ref{fig:collisions/(-1,1)+(1,-1)/examples/epsilon=0.1}.
As discussed previously, for the false vacuum cases, the $(-1, 1)$ kink can still be interpreted as being mostly composed of two subkinks.
The main new feature present in these simulations is the formation of false vacuum bubbles after the annihilation of the inner subkink-antisubkink pair.
These bubbles grow in size as the outer subkinks escape.
However, since the $\phi = 0$ vacuum is false, the bubble requires energy to grow.
At a certain point, the false vacuum bubble stops growing and shrinks back, with outer subkinks moving towards one another and eventually colliding.
After the subkink collisions, there are several possible outcomes.
The entire structure can be annihilated and radiate away ($n = 3$ with $v=0.5$, for example), a new bubble can form ($n = 0.5$ with $v = 0.4$, for example), or the original $(-1, 1) + (1, -1)$ kink-antikink pair can reemerge and escape ($n = 0.5$ with $v = 0.6$, for example).

\begin{figure}
    \includegraphics{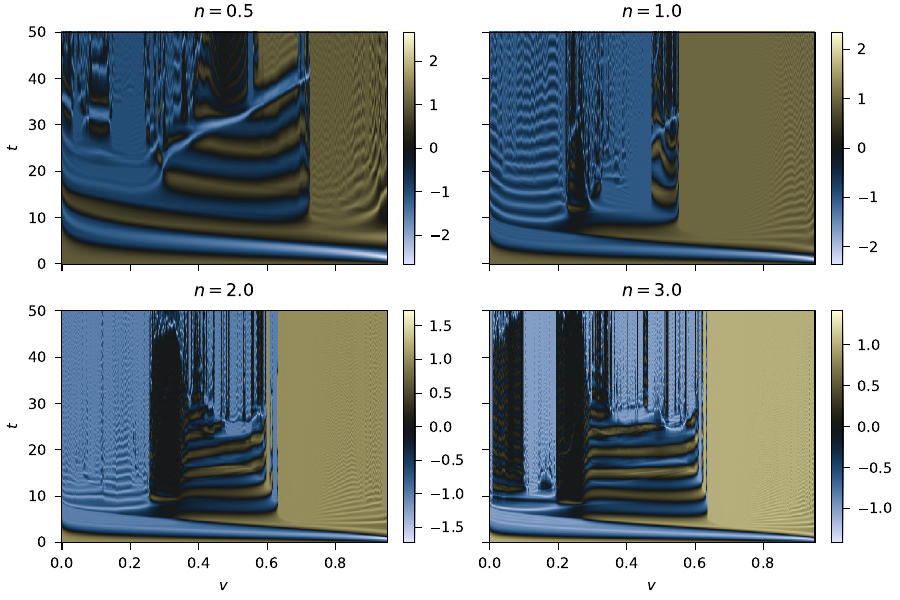}
    \caption{Field $\phi(t, 0)$ for $(-1, 1) + (1, -1)$ collisions as function of velocity $v$ for selected values of $n$ in the case $\epsilon = 0.1$, for which the potential has a false vacuum.}
    \label{fig:collisions/(-1,1)+(1,-1)/middle_examples/epsilon=0.1}
\end{figure}

By looking at the field at the origin as a function of $v$ and $t$ in figure~\ref{fig:collisions/(-1,1)+(1,-1)/middle_examples/epsilon=0.1}, we see that the reappearance of the $(-1, 1) + (1, -1)$ is present only in the $n = 0.5$ case for the time interval simulated ($0 \leq t \leq 50$).
The bubbles in $n = 0.5$ are also special by mostly appearing together with a large central oscillon.
For other values of $n$, we also observe large central oscillons, but bubbles without them are also common (black regions in figure~\ref{fig:collisions/(-1,1)+(1,-1)/middle_examples/epsilon=0.1}).

\begin{figure}
    \includegraphics{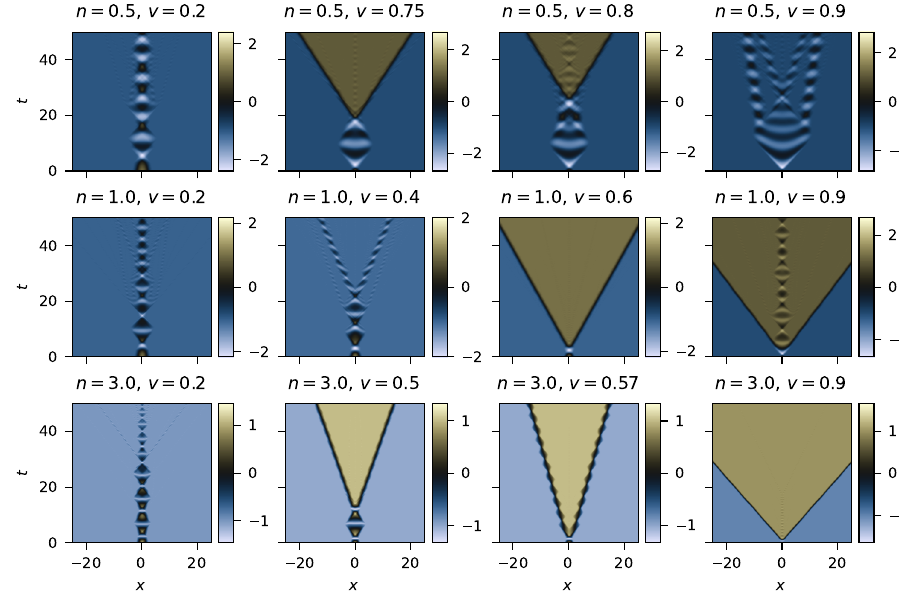}
    \caption{Field $\phi(t, x)$ for $(-1, 1) + (1, -1)$ collisions with $\epsilon = \infty$ and selected values of $n$ and $v$.}
    \label{fig:collisions/(-1,1)+(1,-1)/examples/epsilon=Inf}
\end{figure}

By taking the limit of $\epsilon$ towards infinity, we reach the case for which the potential is a double well.
Some examples of collisions in this case are presented in figure~\ref{fig:collisions/(-1,1)+(1,-1)/examples/epsilon=Inf}.
Some interesting findings are the presence of cases where the pair annihilates, bounces one or more times, and then escapes towards infinity.
Another feature to note is the absence of straightforward escape cases for $n = 0.5$, which are consistent with the results from ref.~\cite{Karpisek:2024zdj} for collisions in a model where the potential is non-analytic at the maxima.
These features can also be observed in the scan of $\phi(t, 0)$ as function of $v$ presented in figure~\ref{fig:collisions/(-1,1)+(1,-1)/middle_examples/epsilon=Inf}.

\begin{figure}
    \includegraphics{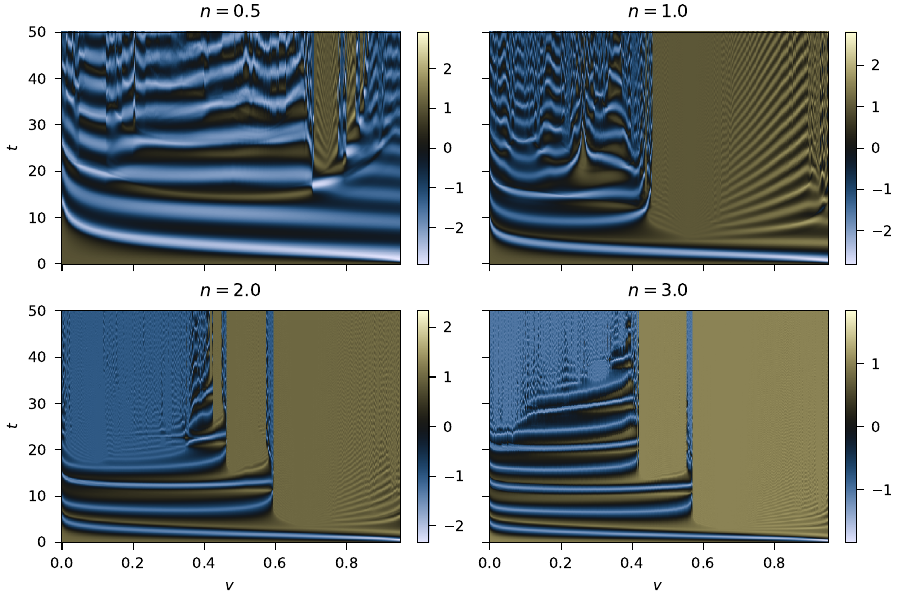}
    \caption{Field $\phi(t, 0)$ for $(-1, 1) + (1, -1)$ collisions as function of velocity $v$ for selected values of $n$ in the case $\epsilon = \infty$, for which the potential has a double well shape.}
    \label{fig:collisions/(-1,1)+(1,-1)/middle_examples/epsilon=Inf}
\end{figure}

\begin{figure}
    \includegraphics{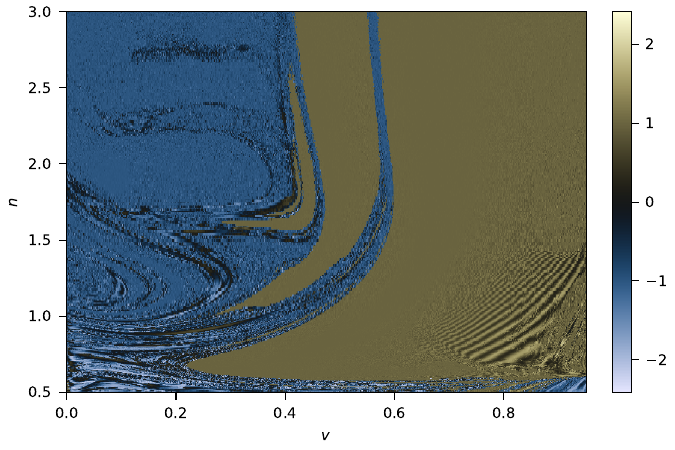}
    \caption{Field $\phi(50, 0)$ for $(-1, 1) + (1, -1)$ collisions as function of velocity $v$ and $n$ in the case $\epsilon = \infty$, for which the potential has a double well shape.}
    \label{fig:collisions/(-1,1)+(1,-1)/middle_final/epsilon=Inf}
\end{figure}

We may also analyse the dependence on $n$ by considering the late time field $\phi(50, 0)$ in figure~\ref{fig:collisions/(-1,1)+(1,-1)/middle_final/epsilon=Inf}.
Looking at the lower region of the plot, we see that only for $n$ slightly larger than $0.5$ we have capture for all velocities.
For larger values of $n$, there is a critical velocity separating the escape cases, as well as one or two windows of bounces followed by escape.

\begin{figure}
    \includegraphics{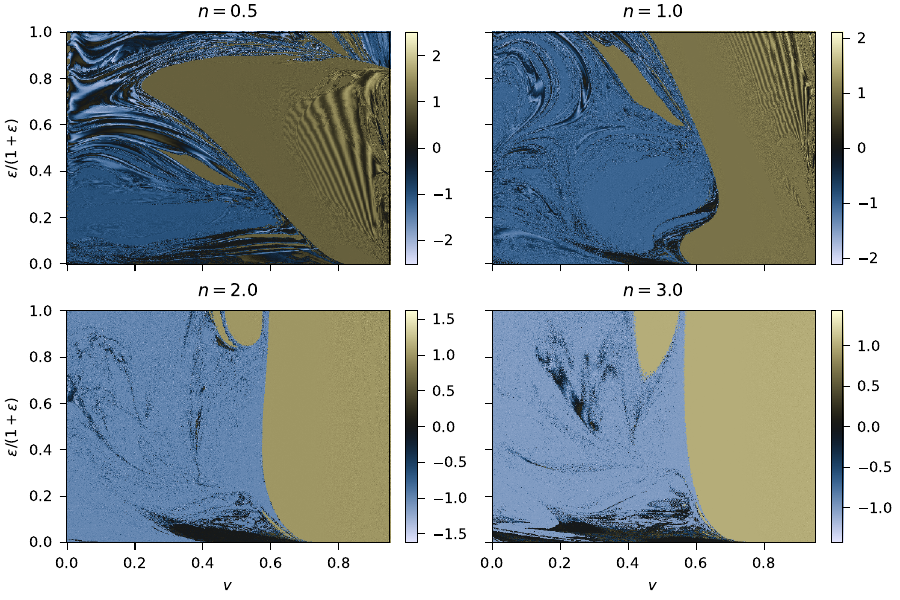}
    \caption{Field $\phi(50, 0)$ for $(-1, 1) + (1, -1)$ collisions as function of velocity $v$ and $\epsilon$ for selected values of $n$.}
    \label{fig:collisions/(-1,1)+(1,-1)/middle_final_by_epsilon}
\end{figure}

At last, we look at the late time field $\phi(50, 0)$ as function of $\epsilon$ and $v$ in figure~\ref{fig:collisions/(-1,1)+(1,-1)/middle_final_by_epsilon}.
We see that capture for almost all velocities is restricted to high values of $\epsilon$ in the $n = 0.5$.
For other values of $n$, we similarly see that the presence of windows of bounce followed by escape are also limited to high enough values of $\epsilon$.

\section{Conclusions}
\label{sec:conclusions}

We studied compact kinks in a modified Christ-Lee model, in which the potential is a non-analytic function at the minima.
The model possesses two parameters, $n$ and $\epsilon$, characterizing the order and vacuum structure of the potential.
We have classified and discussed in detail the vacuum structure in all possible instances, comprising a triple well potential with vacua for field values $\phi \in \{-1, 0, 1\}$, cases with false vacuum $\phi = 0$ and two true vacua at $\phi = \pm 1$, and cases with a double well with vacua at $\phi = \pm 1$.

For all mentioned cases, static solutions for compact kinks have been obtained numerically.
We have found that for the cases of small $\epsilon$, the kink in the $(-1, 1)$ sector can be interpreted as composed of two subkinks in the $(-1, 0)$ and $(0, 1)$ sectors.
We have also analyzed the dynamics of such kinks under small perturbations by finding its internal modes.
As usual in the case of compact kinks, the stability potential for small oscillatory perturbations features infinite barriers at the borders of the compact kinks, resulting in the internal modes to also be compact.
For $\epsilon = 0$, we have found that the presence of the subkinks makes the modes' frequency spectrum degenerate.
However, the degeneracy is quickly broken when increasing values of $\epsilon$.

The interactions between the compact kinks have also been studied through a careful numerical treatment of the compactons collision.
We have found that the parameter $n$ is related to the emission of radiation in kink-kink collisions in the $(-1,0) + (0,1)$ sector, with larger amounts of radiation emitted when $n$ is large, while the $n = 1/2$ case has no radiation.
In the kink-antikink collisions in the $(0, 1) + (1, 0)$ and $(1, 0) + (0, 1)$ sectors, the parameter $n$ has greatly influenced the critical velocity separating the capture and escape cases, as well as the formation of large central oscillons and the size of the oscillons that compose the radiation.

In the case of $(-1, 1) + (1, -1)$ collisions, we have showed that, for small values of $\epsilon$, the internal structure of the kink as two subkinks plays an important role.
Indeed, for $\epsilon$ small, we have detected large windows of velocities for which the subkinks decouple from the larger kink, leading to the formation of false vacuum bubbles.
Even the cases without decoupling, the more rich inner structure of the $(-1, 1)$ kink renders the collision dynamics more complicated, causing the appearance of extra capture windows in the $n = 1/2$, $\epsilon = 0$ case, as well as a window of multi-bounce followed by escape in the $n=1/2$, $\epsilon = 0.1$ case.

Perhaps somewhat surprisingly, also in the simpler double well case ($\epsilon \rightarrow \infty$), we have found that the kink-antikink collision dynamics is less straightforward than previously thought.
The $n$ parameter has been shown to greatly influence the dynamics.
For example, the $n = 1/2$ case displays almost no escape cases, except for a small velocities window, for which escape occurs only after a few bounces.
Cases of multi-bounce followed by escape mostly disappear when we increase $n$ to $1$, but reappear for larger values of $n$.

Our results show that the compact solutions are still far from being fully understood, specially as it relates to their interactions.
Several previous results have been found in this work to be characteristics of specific choices of potential parameters, instead of general properties of this class of solutions.
Further investigation is required to determine if other previous results about compact kinks hold for our more general potential.

In particular, the presence of fractal-like dependence on the collision velocity $v$ needs to be verified.
Another important question is related to the applicability of the collective coordinates approach to describe the interaction of compactons.
The collective coordinates method has found reasonable success in describing the dynamics of kink-antikink collisions in the case of parabolic potentials, both in the double well and periodic cases.
It is natural to wonder whether that method would find the same success in the case of higher order potentials.
Providing answers to such questions is an important topic for future work and requires further challenging theoretical and numerical work.

\begin{acknowledgments}
    We thank P.~Klimas and J.~S.~Streibel for helpful comments.
    This study was financed in part by the Coordenação de Aperfeiçoamento de Pessoal de Nível Superior -- Brasil (CAPES) -- Finance Code 001.
    The authors would like to thank the Centro de Computação Avançada e Multidisciplinar from Universidade Estadual de Santa Cruz for their cooperation, provision and operation of the computing facilities, and their commitment to investing in research, development and innovation for education, industry and society as a whole.
\end{acknowledgments}

\appendix
\section{Numerical methods}

The simulations of the full field equation were performed by discretizing the spatial derivatives with fourth order central finite differences.
The time evolution was performed with the Bogacki-Shampine 5/4 Runge-Kutta method~\cite{Bogacki:1996} through the Julia~\cite{Bezanson:2014pyv} library DifferentialEquations.jl~\cite{Rackauckas:2017}.
The time steps were fixed as $\Delta t = \Delta x / 5$, where $\Delta x$ is the spatial discretization step size.
We have also tested several other integration methods, which yielded consistent results.
For the field and energy density plots of specific velocities, we have used spatial steps of size $\Delta x = 0.001$ and performed the simulations for $x \in [-t_\text{max}, t_\text{max}]$, where $t_\text{max}$ is the final time of the simulation.
Since the collision begins at $t = 0$ with the kinks interacting initially only at $x = 0$, this guarantees that the borders do not influence the results.

For the scans of multiple simulations with $\epsilon = 0$ and $\epsilon = \infty$, such as the ones in figures~\ref{fig:collisions/(0,1)+(1,0)/middle_final} and~\ref{fig:collisions/(1,0)+(0,1)/middle_final}, we have adopted a different strategy.
Since the scans only use the field at $x=0$, we have performed the simulations only for $x \in [-x_\text{max}, x_\text{max}]$, with $x_\text{max}$ chosen as $12$ for the $(1, 0) + (0, 1)$ and $(0, 1) + (1, 0)$ cases, and as $15$ for the $(-1, 1) + (1, -1)$ case with $\epsilon \in \{ 0, \infty \}$.
To guarantee that the borders would not play a role in the simulation outcome, we have introduced a dissipative term to the field equation at a region of size $l_D$ before each border.
That means that, for $x \in [x_\text{max} - l_D, x_\text{max}]$ and $x \in [-x_\text{max}, -x_\text{max} + l_D]$, the actual simulation was performed with the field equation
\begin{equation*}
    \partial_t^2\phi = \partial_x^2 \phi - V_{n,\epsilon}'(\phi) - \alpha_D \, \partial_t \phi.
\end{equation*}
We have chosen $l_D = 2$ and $\alpha_D = 2$ for the $(1, 0) + (0, 1)$ and $(0, 1) + (1, 0)$ cases, and $l_D = 5$ and $\alpha_D = 0.5$ for the $(-1, 1) + (1, -1)$ case, with $\epsilon \in \{ 0, \infty \}$.
For the $(-1, 1) + (1, -1)$ case with $0 < \epsilon < \infty$, we have performed the simulation in the full interval without using the dissipative term due to the formation of large false vacuum bubbles.

The introduction of the dissipative term near the borders guarantees that outgoing structures will not be reflected back to the region of interest.
The smaller grid also meant that we had fewer degrees of freedom that could possibly accumulate numerical error, allowing us to use step sizes as large as $\Delta x = 0.02$.
The introduction of the dissipative borders and the increased step sizes did not change the scan results while severely reducing the required computation time.
The dissipative regions were found to be very effective. The radiation energy  reaching it quickly vanishes, while the solitons stop moving before reaching the borders of the simulated region.
We further improved the computational performance by exploring the symmetry of the initial conditions and only simulating the region $x \geq 0$.

The chosen numerical setup, actually similar to the one used in ref.~\cite{Karpisek:2024zdj}, has shown the best performance among other ones with comparable accuracy.
We controlled the accuracy by simulating standard configurations, such as uniformly moving kinks and signum-Gordon oscillons, and comparing the results with the known solutions.
Despite the non-analytic nature of the potential, in all tested cases, the simulation results reproduced the known solutions within the method accuracy.
Note that the non-analytic potential can be seen as the limit of infinite mass of a more usual potential~\cite{Bazeia:2014hja,Hahne:2024qby}, and the compact kinks are the limiting case of kinks with an increasingly fast approach to the vacuum.
Therefore, any theoretical constraint on the application of the numerical methods could be circumvented by working with a smoothed out potential with high enough mass.
We also checked energy conservation for the cases without the dissipative term and the total energy of the system did not deviate from the initial amount by more than 5\%.

\bibliography{bibliography}

\end{document}